\documentclass{SNmult}

\usepackage[
  left=3cm,
  right=3cm,
  top=4cm,
  bottom=4cm
]{geometry}

\usepackage[T1]{fontenc}
\usepackage{type1cm}
\usepackage{graphicx}
\usepackage[bottom]{footmisc}
\usepackage{newtxtext}
\usepackage[varvw]{newtxmath}

\usepackage{xurl}
\usepackage{placeins}
\usepackage{float}

\usepackage{siunitx}
\usepackage[ruled,vlined,linesnumbered]{algorithm2e}

\begin{document}

\title*{Building an Adversarial Malware Dataset by Family and Type: Generation, Evasion, and Poisoning Evaluation}
\titlerunning{Building an Adversarial Malware Dataset}

\author{David Ko\v{s}\v{t}\'{a}l \and Martin Jure\v{c}ek \orcidID{0000-0002-6546-8953}}
\authorrunning{Ko\v{s}\v{t}\'{a}l, D. and Jure\v{c}ek, M.}

\institute{David Ko\v{s}\v{t}\'{a}l \at Department of Information Security, Faculty of Information Technology, Czech Technical University in Prague, \email{kostada2@fit.cvut.cz}
\and
Martin Jure\v{c}ek \at Department of Information Security, Faculty of Information Technology, Czech Technical University in Prague, \email{martin.jurecek@fit.cvut.cz}}

\maketitle

\abstract{We present a dataset of adversarial malware samples derived from the public RawMal-TF collection of real-world malware binaries. Using a suite of adversarial malware generators, we construct two sets of adversarial PE files: 44,347 family-labelled samples and 33,596 type-labelled samples, achieving evasion rates of 98.35\% and 92.20\% against the EMBER classifier, respectively. Each adversarial binary is accompanied by detailed metadata, including EMBER scores and VirusTotal classifications. We further demonstrate the susceptibility of malware classification pipelines to data poisoning attacks through a series of training experiments. Injecting fully mislabelled adversarial samples representing only 0.5\% of the training data in the family-labelled dataset increases the evasion rate against the re-trained classifier from 26.1\% to 92.8\%. The dataset is publicly released to facilitate future research on adversarial malware, poisoning attacks, and the robustness of machine-learning-based malware detection systems.
}

\section{Introduction}\label{sec:intro}

The field of malware detection and cybersecurity more broadly is ever-evolving. Adversaries constantly devise new attacks, making it difficult for defenders to keep pace. Malware classifiers based on machine learning (ML) have been used for over two decades in the malware arms race \cite{schultz}, as they enable cybersecurity professionals to detect emerging threats faster. Traditional classification techniques relying solely on static signatures are no longer sufficient \cite{moser}. As the role of machine learning in malware detection increases, malicious actors have a growing incentive to evade ML-based malware detection software \cite{grosse}. This naturally spurs new research in the malware community, aiming at enhancing defences.

The challenge of adversarial attacks in machine learning is well known and difficult to fully address. Adversarial attacks in ML involve intentionally altering classifier input data in a way that causes misclassification. In the context of malware, adversarial (evasive) samples are malicious executable files, modified to appear benign when classified by an ML detector, while still preserving their original functionality.

The goal of this work is to create a dataset of adversarial malware samples from real-world malware, based on the existing open RawMal-TF collection \cite{rawmal}. Such a dataset can be used to enhance the resilience of ML-based classifiers against adversarial attacks, for example by incorporating modified samples into the training phase.

Existing adversarial malware generators \cite{comp} typically demand complex setup and significant compute resources, as iterative binary modification paired with continuous classification does not scale easily. Accessible datasets of real-world raw malware are similarly scarce. To our knowledge, no resource provides binary adversarial samples created from real malware at scale, leaving a gap this work directly addresses.

We also perform data poisoning experiments: we inject samples from our newly created adversarial dataset into malware classifier training sets in different configurations, and observe how metrics change.

In this work, we are only considering Portable Executable (PE) binaries, used by the Microsoft Windows operating system. This is by far the most researched format in the malware space, as Windows has been the dominant target for both adversaries and researchers. To generate adversarial samples at scale, we use automatic adversarial malware generators, which take malware samples as input, attempt to perform functionality-preserving modifications, and test whether the resulting sample has successfully evaded classification.

The remainder of this work is organized as follows. Section 2 introduces the necessary background on the base dataset and the EMBER classifier. Section 3 presents adversarial malware generators used in this work. Section 4 describes the dataset creation pipeline and experimental environment. Section 5 analyses the generated samples in terms of evasion performance, antivirus detection, file size overhead, and final dataset composition. Section 6 evaluates the impact of data poisoning attacks using the proposed dataset, and Section 7 concludes the paper and summarizes the achieved results.

\section{Background}\label{sec:bg}

This section aims to explain concepts relevant to the topic of this work. This includes an overview of our base dataset and the malware classifiers we utilise in our work.

\subsection{RawMal-TF dataset}\label{subsec:rawmal}
RawMal-TF \cite{rawmal} is a dataset of real-world raw malware files, split into two groups: samples labelled by type, and samples labelled by family. This work heavily relies on this dataset; it serves as a basis for all of our adversarial samples. The dataset is constructed from public sources: VirusShare, vx-underground and MalwareBazaar\footnote{Online repositories hosting malware samples, available at \url{https://virusshare.com}, \\ \url{https://vx-underground.org} and \url{https://bazaar.abuse.ch}, respectively.}. The family-labelled group contains 17 distinct families, while the type-labelled group features 14 types.

\subsubsection{Dataset construction}
Files for the type-labelled dataset originate from VirusShare. Type labels (trojan, spyware, virus, dropper etc.) were assigned using ClarAVy \cite{claravy}, a tool for parsing VirusTotal~(VT) \cite{virustotal} results. ClarAVy assigns labels based on classifications from a large number of antivirus (AV) vendors.

To achieve that, ClarAVy splits AV classification labels in various formats into unified tokens and performs various transformations (e.g., merging equivalent classes). To select final labels, ClarAVy employs statistical methods. The full process is described in the ClarAVy paper.

Files without behavioural (i.e., type) labels or categories that have fewer than 1,000 samples were excluded from the type-labelled dataset. RawMal-TF uses precomputed ClarAVy labels for VirusShare samples, distributed via vx-underground.

Samples for the family-labelled dataset were obtained from vx-underground, in the form of archives originating from MalwareBazaar. Names of those samples include their families, so the authors assigned labels by systematically parsing filenames. Files with unclear or generic labels were excluded. Families with fewer than 1,000 samples are also excluded from the final family-labelled dataset.

RawMal-TF also provides per-family and per-type feature sets, directly compatible with the original EMBER format \cite{ember}. Feature files contain both malicious and benign samples.

\subsection{EMBER}\label{subsec:ember}
EMBER \cite{ember}, published in 2018, is a feature format and a dataset of features extracted from PE files. It includes malicious, benign, and unlabelled samples, consisting of 1.1M files in total (in its original release). The authors also provide an extraction pipeline and  classification models. EMBER is widely used in many academic works in this field  as a benchmark and a standard feature definition. It extracts the following features:
\begin{itemize}
\item general file information such as file size, virtual size, import and export counts,
\item flags for whether debug sections, TLS, resources, relocations, or signatures are present,
\item header fields including target machine, image characteristics, subsystem, DLL characteristics, magic, and version numbers,
\item imported libraries and functions as well as exported functions,
\item section properties including name, size, entropy, and characteristics,
\item a histogram of byte value distributions across the file,
\item a joint histogram of entropy and byte values over sliding windows,
\item string statistics including count, average length, character distribution, entropy, and counts of paths, URLs, registry keys, and `MZ' (PE file) signatures.
\end{itemize}

EMBER provides multiple versions of their dataset: samples from 2017  with version~1 features, samples from 2017 with version 2 features, and samples from 2018 with version 2 features. The classification models utilised by EMBER are using gradient-boosted decision trees (GBDT), implemented with the LightGBM library\footnote{Available from \url{https://github.com/lightgbm-org/LightGBM}.}. The authors claim that their original model (trained on their 2017 dataset) achieves less than $0.1\%$ false positive rate (FPR) at a detection rate exceeding $92.99\%$. At less than 1\% FPR, the model exceeds $98.2\%$ detection rate.

This work uses a classifier trained on the EMBER 2018 dataset.\footnote{More precisely, the EMBER 2018 classifier from the MAB-Malware Docker image: \url{https://hub.docker.com/r/wsong008/mab-malware}.}
When referring to EMBER as a classifier in this work, we mean this model.

\subsubsection{EMBER2024}
EMBER2024 \cite{ember24}, published in 2025, extends EMBER with features extracted from 3.2M labelled samples, collected between September 2023 and December 2024 across multiple file formats. PE files make up the majority of the dataset, with APK, ELF and PDF samples also included. Many samples in the dataset are also labelled with family, behaviour, packer, exploited vulnerability and threat group labels. EMBER2024 also introduces a \textit{challenge} set of 6,315 malicious files that initially evaded all VirusTotal antivirus products. The new accompanying feature format, sometimes called EMBERv3, retains the categories from the original paper and adds:

\begin{itemize}
\item extended PE header, optional PE header and PE section header fields,
\item PE data directory entries and Authenticode signature properties,
\item Rich header (undocumented proprietary metadata from Microsoft's linker),
\item PE parse warnings and errors,
\item expanded string pattern matching (76 patterns).
\end{itemize}

EMBER2024 also supports limited extraction from non-PE files. The released LightGBM detection model achieves a detection rate of $94.48\%$ at $1\%$ FPR on the test set, with performance dropping substantially on the challenge set.

\section{Adversarial malware}\label{sec:adv}

Adversarial examples are inputs crafted to mislead ML models while remaining valid in some meaningful sense. In the malware domain, this means modifying an executable so that it retains its malicious functionality yet evades a classifier. Formally, given a malware sample $X$, a functionality-preserving transformation $G$ produces an adversarial sample $X' = G(X) = X + \delta$, where $\delta$ is the perturbation introduced by $G$.

Such transformations can be performed in numerous ways \cite{amg}, including by hand. However, manually crafting adversarial samples does not scale to a dataset of any meaningful size, so we instead rely on automated generation. For this purpose, we use adversarial malware generators, which we examine in the following section.

\subsection{Adversarial malware generators}

Adversarial malware generators are software tools which automatically perform functionality-preserving modifications (perturbations) on malicious executables, with the goal of evading detection by ML models. Generators can target either a white-box or a black-box setting.

White-box generators rely on knowing the internal state of the targeted ML model. That may help with making more efficient modifications to binary files, as the generator has more insight into what causes the model to classify a sample in a certain way. However, in the real world (e.g., when targeting a commercial AV product) it is uncommon to have access to model internals.

In such situations, black-box generators are more useful. These rely solely on the binary output of the target classifier: whether the sample is classified as benign or malware, with no access to the internal model state. Models trained in a black-box setting also tend to generalise better and may transfer more readily to other classifiers. However, black-box generators are typically slower to achieve evasion and may exhibit a lower overall success rate compared to their white-box counterparts.

Adversarial malware generators employ a wide range of techniques \cite{comp}. White-box generators may use gradient-based attacks, while black-box generators may utilise reinforcement learning (RL), evolutionary algorithms or other adaptive algorithms. A simpler generator could also simply apply random mutations. There is a variety of off-the-shelf academic adversarial malware generators. This section provides an overview of such projects.

\vspace{-0.25cm}

\subsubsection{Gym-malware}
Gym-malware \cite{gym} is a pioneering RL-based black-box generator, which inspired much of the subsequent research. This work introduced the concept of using RL techniques in adversarial malware generation (i.e., rewarding the agent for performing modifications that successfully evade a given classifier). It also features the following modifications, many of which are shared by other generators:

\begin{itemize}
\item adding an unused entry to the import table,
\item manipulating existing section names or adding new unused sections,
\item appending bytes to extra space at the end of sections or end of the file,
\item creating a new entry point (immediately jumping to the original one),
\item removing signing information or breaking the PE checksum,
\item manipulating debug info,
\item UPX packing or unpacking the file\footnote{Refers to UPX: the Ultimate Packer for eXecutables, \url{https://upx.github.io/}}.
\end{itemize}

Authors of this research also provide an environment for the OpenAI Gym framework (now Gymnasium) \cite{gymoai}, a framework for developing RL algorithms. This allows other work to expand on this project.

In the authors' testing on 200 holdout samples across 4 malware families, the RL agent evaded the classifier between 10\% and 24\% of the time depending on the dataset, outperforming the random policy by approximately 2 percentage points. Cross-testing on VirusTotal showed a median reduction of more than 50\% in the number of AVs flagging the samples as malicious. However, the results also revealed a limitation: on the VirusShare dataset, the trained agent reduced the median detection count from 54/65 to 26/65, while the random agent achieved an even lower count of 16/65, suggesting that the learned policy did not transfer well to real-world AV products. Across all datasets, the most consistently effective modification was UPX packing.

\vspace{-0.25cm}

\subsubsection{MAB-Malware}
MAB-Malware \cite{mabm} is a black-box, RL-based generator. Authors claim over 74\%--97\% evasion rate for open-source ML detectors and over 32\%--48\% evasion rate for commercial AVs deployed on VirusTotal. This generator consists of two components: Binary Rewriter and Action Minimizer. Binary Rewriter has a set of mutations, which it attempts to apply until the malware sample evades detection. These modifications consist of:

\begin{itemize}
\item appending benign content to an unused part of a section,
\item adding a new section with benign content,
\item adding benign content to the file overlay,
\item replacing certain instructions with semantically identical ones,
\item adding a single byte to the overlay or an unused part of a section,
\item adding a 1 byte long section or changing 1 byte in an existing section's name,
\item wiping the code signature, checksum or debug symbols in the PE header.
\end{itemize}

The Action Minimizer component will then attempt to remove modifications, such that the binary is still evading detection, to obtain a minimal sample and to help understand which modifications work the best and reduce the chance of breaking original functionality. MAB-Malware supports multiple targets: open-source ML-based classifiers EMBER and MalConv \cite{malconv}, or a commercial antivirus in a virtual machine.

\subsubsection{Pesidious}
Pesidious \cite{pesidious} is a black-box adversarial malware generator that uses generative adversarial networks (GANs) and RL, and it has an included malware classifier (a model from Gym-malware). Its GAN component is used for generating benign-looking import tables and sections, while its RL model attempts to choose the best sequence of mutations, including the mutations generated by the GAN. These modifications consist of:
\begin{itemize}
\item adding imports or sections (generated by the GAN model),
\item appending bytes to sections or renaming them,
\item UPX packing or unpacking,
\item adding or removing a signature,
\item appending a random number of bytes to the file.
\end{itemize}

Pesidious is primarily an open-source proof-of-concept and does not accompany a formal evaluation paper. As such, no benchmark results are provided by the authors.

\subsubsection{SecML Malware}
SecML Malware \cite{secmlm} is a Python library designed to help implement and test adversarial attacks on PE binaries. It also includes a variety of already implemented black-box and white-box attacks utilising different techniques.

GAMMA (Genetic Adversarial Machine learning Malware Attack) \cite{gamma} is a black-box technique for generating adversarial binaries. It consists of two separate components: \textit{section injection} attack and \textit{padding} attack. GAMMA Section injection is one of the more effective attacks included in the SecML package \cite{comp}.

As hinted at by its name, this attack is based on a genetic algorithm, trying to find the best modification by ranking and combining different sets of perturbations. The padding attack consists of adding extra bytes at the end of the file, while section injection adds entire sections and modifies the section table of the PE file.

The authors' testing with a dataset of 200 samples on VirusTotal shows that $46.56$ ($\pm 12.40$) average detections of original samples were reduced to 34.50 ($\pm 12.63$) with the section injection attack.

\subsubsection{AMG}
AMG (Adversarial Malware Generator) \cite{amg} is a framework focusing specifically on generating valid (fully functional) samples. AMG is a black-box attack and employs an RL algorithm to modify binaries. It can perform the following modifications:
\begin{itemize}
\item removing debug symbols or an Authenticode signing certificate,
\item adding a new section or appending a random item to the import table,
\item appending benign data at the end of a section or end of the file,
\item incrementing or decrementing a timestamp in the PE header,
\item renaming a section or zeroing out the PE checksum.
\end{itemize}

In authors' testing against the EMBER classifier, the PPO-based AMG agent achieved an evasion rate of 53.84\%, outperforming other tested RL algorithms. However, the agent did not transfer well to other classifiers, dropping to 11.41\% against MalConv and 2.31\% on average against commercial AVs on VirusTotal. Interestingly, the random agent achieved a comparable 11.65\% against AVs, suggesting that the learned policy provides limited benefit beyond the specific target classifier.

\subsubsection{MalwareTotal}
MalwareTotal \cite{mt} is a more recent framework for bypassing ML detectors. It employs PPO to learn how to modify the binary in the most effective way. In authors' testing, MalwareTotal exhibits significantly greater evasion rates in similar or lesser time compared to GAMMA Sections and MAB-Malware, especially against selected top-rated commercial AVs.  MalwareTotal has support for most modifications out of all tested adversarial generators:
\begin{itemize}
\item extending and injecting random bytes to the DOS stub,
\item zeroing out the PE checksum or debug directory entry,
\item randomising the linker and OS version fields in the optional header,
\item wiping the Authenticode signing certificate from the optional header,
\item renaming a section using a name drawn from benign binaries,
\item adding an unused library import or shifting content before the first section,
\item counterfeiting the import address and name tables,
\item injecting content extracted from benign binaries to a new section or to unused space within an existing section,
\item appending random bytes or extracted benign content to the overlay,
\item modifying bytes across the DOS header, optional header, and section table,
\item packing the binary (instruction reordering / code randomisation).
\end{itemize}
The authors claim MalwareTotal's evasion rate against EMBER is 98.67\% compared to GAMMA Sections' 74.26\% and MAB-Malware's 76.52\% in a comparable time. MalwareTotal also evades commercial AV solutions AMS A and AMS B with rates of 92.63\% and 98.52\% and in lesser time, compared to GAMMA Sections' 17.33\% and 14.93\% respectively. Although MalwareTotal's source code is not publicly available, the authors kindly provided access upon request.

\vspace{1cm}
\section{Dataset creation}\label{sec:ds}

The dataset creation process involved the following high-level steps:

\begin{itemize}
\item researching and selecting adversarial malware generators to evaluate,
\item deploying adversarial malware generators on a faculty-provided server,
\item modifying various generators to fix bugs and to evaluate different parameters,
\item designing and writing scripts to launch various generators in parallel,
\item writing scripts to classify the original dataset and modified samples,
\item evaluating which generators provide suitable results on a reduced dataset,
\item running generators, collecting suitable samples and evaluating results.
\end{itemize}

This section aims to explain how we approached and executed these tasks. The results of the research phase are summarised in the previous section, which briefly describes each generator which we tested. Naturally, there are more generators and methods to consider. However, they were all considered insufficiently effective from the start, had significant functionality overlap with other existing generators, or were inaccessible to the public. We exclude Gym-malware from further sections of this work, as it overlaps substantially with MAB-Malware with its similar reinforcement learning approach and action space.

We are creating a dataset using a suite of five adversarial generators (Pesidious, MAB-Malware, AMG, GAMMA Sections from SecML Malware and MalwareTotal) across six configurations; MalwareTotal is used both with EMBER and with EMBER2024 as its target classifiers. Our final dataset will contain at most one adversarial sample per source sample: if multiple generators create an adversarial sample, we choose the best one using a process described in Sect.~\ref{sec:evals}. If all generators fail to generate a sample, we omit that source sample from the dataset.

We are using EMBER to evaluate the evasiveness of adversarial samples, and three out of six generators configurations use EMBER as their target classifier. However, a sample is included in the final dataset even if it was not originally classified as malicious by EMBER. Likewise, it is included even if the adversarial variant is not classified as malicious. This ensures greater sample diversity, as EMBER has relatively low recall on the source datasets (42.1\% for family-labelled, 73.4\% for type-labelled). We also include samples for which their EMBER scores increased. Metadata for all samples is provided, including EMBER scores, allowing arbitrary filtering if desired. A complete list of metadata together with download links for the final dataset containing adversarial malware samples categorised by both families and types is available at \url{https://github.com/CS-and-AI/AdvMal-TF}.

\subsection{Deploying adversarial generators}

Before beginning to work on creating the dataset, we first need to deploy selected adversarial malware generators to a faculty-provided server. Some generators also require modifications to work in the desired environment.

All of the adversarial malware generators utilised in this work are implemented primarily in Python. However, they often rely on different and outdated versions of various libraries and the language itself. Therefore, we decided to create Docker images for every generator, enabling them to operate in more reliable and reproducible environments, as well as isolating them. Containerisation helps with easier parallelisation and ensures secure operation, since generators work with live malware.

Making all adversarial malware generators run reliably is a challenging task, since they often have specific dependencies and are not usually designed for reliable production operation. All evaluated generators are chiefly research projects, not robust software packages. During testing, we encountered generator bugs and process hangs. We were able to fix some of these problems. However, generators' codebases are complex, and extensive modifications are beyond the scope of this work. We built tooling to handle process hangs, and either retry or fail gracefully without blocking the whole process.

Our environment, where the generators were deployed, the dataset was generated and the evaluation was performed, consists of the following: Ubuntu 22.04 Jammy Jellyfish Linux OS, running on an NVIDIA DGX Station A100 server, with four NVIDIA A100 40 GB GPUs, an AMD EPYC 7742 64-core (128-thread) CPU and 512 GB of RAM.

\subsubsection{Configuration per generator}

MAB-Malware worked out-of-the-box, since the project provided a ready-made Docker image. No further modifications were necessary. For Pesidious, GAMMA Sections, AMG and MalwareTotal, we needed to create container image definition files (Dockerfiles). Additionally, they all needed adjustments to their dependencies.

For Pesidious, we did not need any further changes. For GAMMA Sections, we had to create a custom script which initialises all libraries, classifies files, then mounts the attack on each executable and saves results. The code is based on an example notebook provided by the project\footnote{Notebook available from \url{https://github.com/pralab/secml_malware/blob/master/blackbox_tutorial.ipynb}.}. For AMG, we are using its random agent mode to enhance the dataset's diversity, as the random agent is more successful at evading commercial AVs (based on the authors' testing).

For MalwareTotal, we first had to modify the codebase to resolve issues preventing the generator from operating in our Linux environment and to enable our mass sample generation use case. That involved changing some library imports, fixing bugs and adjusting features, as the original project was not intended to mass generate adversarial samples in a way this work relies on. Rather, it was chiefly designed for evaluating the adversarial generator's performance.

MalwareTotal's original attack is also using a Windows-only commercial packer for its packing modification, so we had to replace it with a free cross-platform solution. We chose the PEzor \cite{pezor,pezorgh} open-source packer, since it is designed to be Linux-native and features advanced obfuscation. It also includes SGN \cite{sgn}, a binary payload encoder with a minimal stub. Following that, we decided to re-train MalwareTotal's \cite{ppoa} agent for EMBER, as the pre-trained one did not include support for packing. We also ran the generator in a different environment and with a different packer.

MAB-Malware, AMG and MalwareTotal all use EMBER as the target classifier to assess whether samples are evasive. GAMMA Sections uses MalConv \cite{malconv} as its classifier, while Pesidious uses a different model from Gym-malware. Some adversarial generators also allow for different settings. We decided to lean towards their default configurations.

In addition to running adversarial generators against their default classifiers, we modified several generators to use EMBER2024 as their target classifier. EMBER2024 exhibits significantly higher recall on our source datasets compared to EMBER (42.1\% for EMBER vs 97.6\% for EMBER2024 in our source family-labelled dataset). EMBER2024 appears more robust to adversarial modifications in our experiments. The generators evaluated against EMBER2024 achieved a relatively low adversarial sample generation success rate. Nevertheless, we chose to include them in order to increase the diversity of the dataset, as they introduce a broader range of modifications to the malware samples.

After evaluating different generators, we have decided to run MalwareTotal with EMBER2024, apart from just with EMBER. We have modified the codebase to allow the agent to communicate with an HTTP endpoint providing classification scores for input data. The EMBER2024 classifier itself runs in a separate custom Docker container and exposes an HTTP server for the agent. We also trained a new MalwareTotal \cite{ppoa} to support this configuration. We refer to this setup as \textit{MalwareTotal (E24)}.

For generators which require benign executables (goodware) or strings for injecting into the adversarial samples, we used several applications from PortableApps.com and selected executables from the Chromium browser. Strings were also extracted from these binaries. Goodware is used by the GAMMA Sections attack and MalwareTotal, while strings are only used by MalwareTotal.

Note that adversarial generator performance comparison is not a primary goal of this work. Generators are adapted for our use case as explained above, and their results may differ from their optimal settings.

\subsection{Creating tooling}

To facilitate running adversarial malware generators on larger scale in parallel, with improved control and reliability, and to better isolate process hangs and other bugs, we created a custom harness script. Harness automatically splits the input dataset into many chunks, launches worker containers to achieve optimal system load, monitors running containers while displaying their status, and merges outputs after the process finishes. Worker launches are driven by the system load average metric, so system resource use is optimised. All adversarial generators only utilise the server's CPU. GPUs are unused at this stage.

Harness detects containers that are frozen by monitoring their log outputs. We consider logs stale if they have not been updated within a certain timeframe. Containers with stale logs can be restarted a certain amount of times, and if they freeze again, they will be discarded. It also provides a terminal user interface in the form of a table, showing current status for running containers. Before launching workers, harness selects only valid PE files, since the type-labelled dataset also contains other file types.

In the configuration for the final dataset runs, we allow for one worker restart when logs are stale for 10 minutes. If the worker fails to complete its run again, we discard its inputs and outputs. To lower the impact of such failures, the dataset is split into approximately 2,000--4,000 parts, depending on the generator.

To assess and pick samples according to criteria defined in the next section, we created multiple scripts to scan the output folders of generators. Each one uses a different classifier, and every script will create a JSON file with classifications data. Scripts for generating EMBER and EMBER2024 classifications run their classifier in parallel to better utilise system resources.

Script for generating VirusTotal classifications first attempts to load an existing analysis report. In case a sufficiently recent analysis is not available, the sample gets uploaded and the newly created analysis ID is put into a queue. The script then periodically performs status checks for older analyses. Since our VT API requests have a daily limit, the script also supports saving its state to disk. This allows us to successfully complete the analysis, seamlessly relaunching the script multiple times over several days.

A final script in the pipeline consumes JSON files created by the aforementioned classification scripts and copies the best samples into a destination folder. JSON classifications are also used for analyses and visualisations via a custom Jupyter notebook. The algorithm for picking the best samples is described in the following section.

\subsection{Evaluating samples}\label{sec:evals}

To assess the quality of generated samples and select the best ones (i.e., samples with the highest evasion rate), we decided to use scores from the pretrained EMBER (2018) classifier \cite{ember} as our primary evaluation criterion. EMBER has been the long-standing standard in academic research, facilitating easier comparison with prior research. File size increase is also taken into account as a secondary parameter.

As mentioned in the previous section, EMBER is also significantly easier to evade than modern ML classifiers and commercial AVs. Every generator we tried appears to be incapable of evading advanced classifiers within reasonable time and space constraints. To enhance the utility of the dataset, we additionally provide scores from the pretrained EMBER2024 classifier. We also provide a subset of results from VirusTotal \cite{virustotal}.

VirusTotal score evaluation consists of two sections: all available AVs and top 10 AVs, i.e., 10 antiviruses labelled as a top product that we selected from a recent evaluation by AV-TEST \cite{avtest}. In our observations, leading AVs exhibit greater resilience towards adversarial attacks, so we chose this split to get more representative results. Names of antivirus vendors are anonymised, as this work is not intended to compare AV products and AVs deployed on VirusTotal may run with different settings compared to AV products shipped to customers (e.g., without cloud-based dynamic analysis).

This work omits functionality validation of generated samples, as we consider it beyond the scope. All generators we used are designed to use only functionality preserving modifications. Unfortunately, generated binaries can sometimes be faulty regardless, due to limitations of PE manipulation libraries, bugs, or other factors \cite{amg}.

The function for picking best adversarial samples from generator outputs is described in Algorithm~\ref{alg:algo}. The EMBER threshold constant was chosen based on testing in the original work as a safe option that results in an FPR below 0.1\%. However, it should be noted that this threshold was selected based on the performance figures reported for the 2017 dataset variant, whereas this work uses the 2018 model throughout. The threshold was not re-evaluated against the 2018 model, and the exact FPR guarantee may therefore not hold.

\begin{algorithm}[h]

\caption{Function for picking the final samples}\label{alg:algo}

\KwIn{list of \{generator, ember\_score, orig\_size, modified\_size\}}

\KwOut{best\_generator}

EMBER\_THRESHOLD $:=$ 0.871\;

SIZE\_RATIO\_THRESHOLD $:=$ 1.5\;

MAXIMUM\_SIZE $:=$ 25,000,000\;

best\_score $:=$ $\infty$\;

best\_generator $:=$ nil\;

\For{ignore\_size in [false, true]}{

  \For{file in files}{

    \If{file.modified\_size $>$ MAXIMUM\_SIZE}{

      \textbf{continue}\;

    }

    ratio $:=$ file.modified\_size / file.orig\_size\;

    \If{file.ember\_score $<$ best\_score}{

      \If{ignore\_size \textbf{or} ratio $\leq$ SIZE\_RATIO\_THRESHOLD}{

        best\_score $:=$ file.ember\_score\;

        best\_generator $:=$ file.generator\;

      }

    }

  }

  \If{best\_generator $\neq$ nil \textbf{and} best\_score $<$ EMBER\_THRESHOLD} {

      \textbf{break}\;

  }

}

\Return{best\_generator}\;

\end{algorithm}

The size ratio threshold ensures that the algorithm prefers adversarial samples with a modest size increase (set to $1.5 \times$), since some generators tend to inflate binary sizes. The maximum size threshold of 25,000,000 bytes, or 25 MB, was chosen as a reasonable safeguard against bloated adversarial samples. Before the final dataset is constructed, we also drop samples which are not valid PE files.

The EMBER score represents the probability or confidence with which the model classifies a file as malware. A lower score corresponds to a more successful adversarial sample, as it is less likely to be detected by EMBER. During the first pass (\texttt{ignore\_size = false}), the algorithm prefers samples that achieve a low EMBER score while maintaining a reasonable file size increase. If no sufficiently good sample is found, the size constraint is ignored during the second pass (\texttt{ignore\_size = true}).

\subsection{Dataset creation methodology}\label{sec:metho}

To verify reliable operation and to evaluate generator output quality and performance, we first performed testing on a reduced dataset. This dataset was constructed from RawMal-TF family-labelled dataset by picking the first 50 samples for each family (with an earlier version of the dataset containing more families) for a total of 3,500 input files. We decided to proceed with all tested generators, since all of them provide compelling performance in our evaluations and have varying strengths and weaknesses.

Following this step, we ran all generators, one by one, for both sections of the dataset (type-labelled and family-labelled). We were then able to select the best samples with our scripts, according to Algorithm~\ref{alg:algo} explained in the previous section. Unfortunately, we are unable to provide reliable run times for each generator, as our server has multiple users and exhibits variable system load at different times. However, the time efficiency of different generators varies greatly, with MAB-Malware being the fastest, while MalwareTotal (E24) was the slowest.

During the final dataset validation, we discovered that Pesidious manifests two unexpected failure modes, causing invalid adversarial samples to pollute the output dataset. The first failure mode was sample collapse, where the generator returns an identical adversarial sample for multiple different input files (332 occurrences in the family-labelled dataset, 2143 occurrences in the type-labelled dataset). The second failure mode was returning unmodified samples, where the output is identical to the input (59 occurrences in the family-labelled dataset, 443 occurrences in the type-labelled dataset). We mitigated these failures by updating our pipeline to reject these samples from the final dataset, replacing them with alternative ones.

Additionally, during the process of creating the type-labelled dataset, we filtered out anomalous oversized samples generated by AMG, some of which had over 1 gigabyte. Such files were rejected by VirusTotal.

EMBER scores for MAB-Malware, AMG and MalwareTotal need to be considered separately from scores of GAMMA Sections, Pesidious and MalwareTotal~(E24), since the first group uses EMBER as their target classifier, while the second group uses different classifiers. For generators trained specifically for EMBER, high evasion rates did not transfer effectively to commercial AVs, especially for top products. Generators we tested trained for other classifiers generally achieved greater success in evading commercial AVs.

\section{Dataset analysis}\label{sec:analysis}

Not all generators produce equally useful adversarial samples, and raw evasion rate alone is an insufficient measure of generator quality. This section examines the outputs of the generation process from multiple angles: EMBER evasion performance, commercial antivirus detection, and file size overhead, before reporting the final composition of both datasets.

\subsection{Analysing the family-labelled dataset}

\begin{figure}[h]
\includegraphics[width=\textwidth]{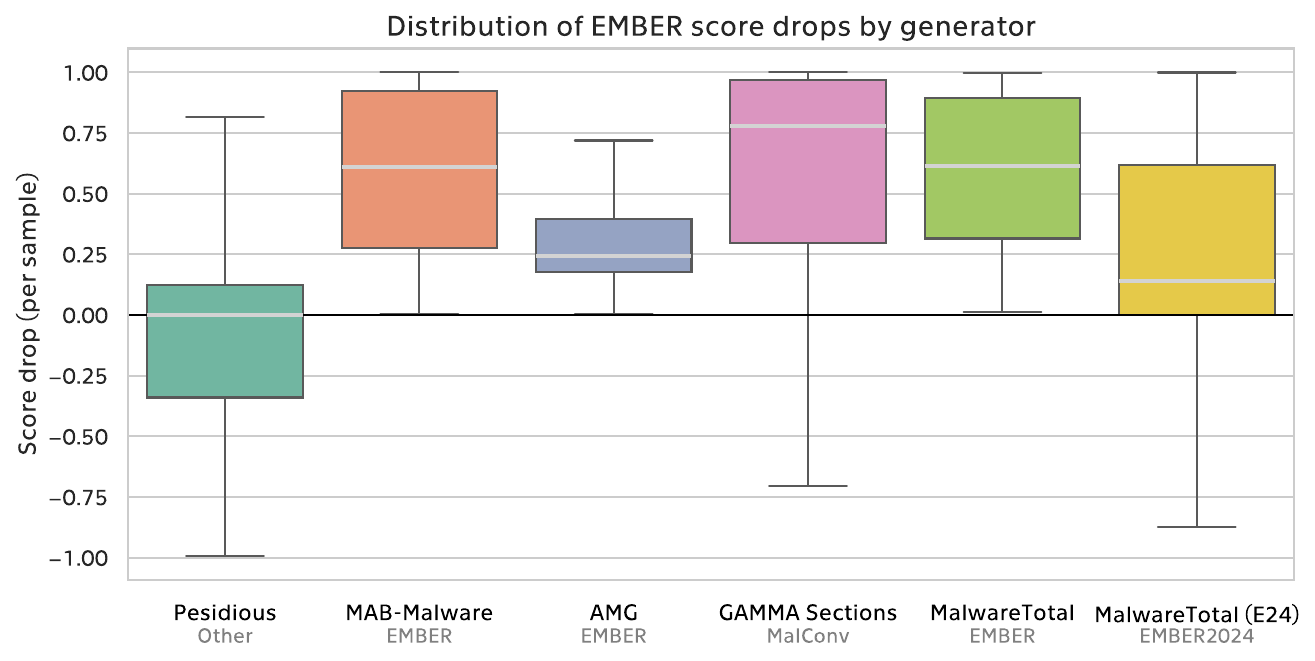}

\caption{Box plot of EMBER score drops by the generator. Note that generators using EMBER as their target never make the score worse, while others might.}
\label{fig:fe}
\end{figure}

Figure~\ref{fig:fe} presents EMBER score drops per generator: unsurprisingly, generators trained for the EMBER target classifier were efficient at lowering EMBER scores. Lower efficiency for AMG can be explained by using its random agent mode, which was not as efficient at evading EMBER. GAMMA Sections was successful across the board. Nevertheless, it also consistently caused samples' file size to grow the most.

Pesidious' Gym-malware classifier flagged a much larger share of source samples as malicious than EMBER did, so Pesidious attempted modification on more inputs. That led to Pesidious generating a large number of samples. Pesidious was not consistently successful at evading EMBER, but it fared better with other classifiers.

\begin{figure}[h]
\includegraphics[width=\textwidth]{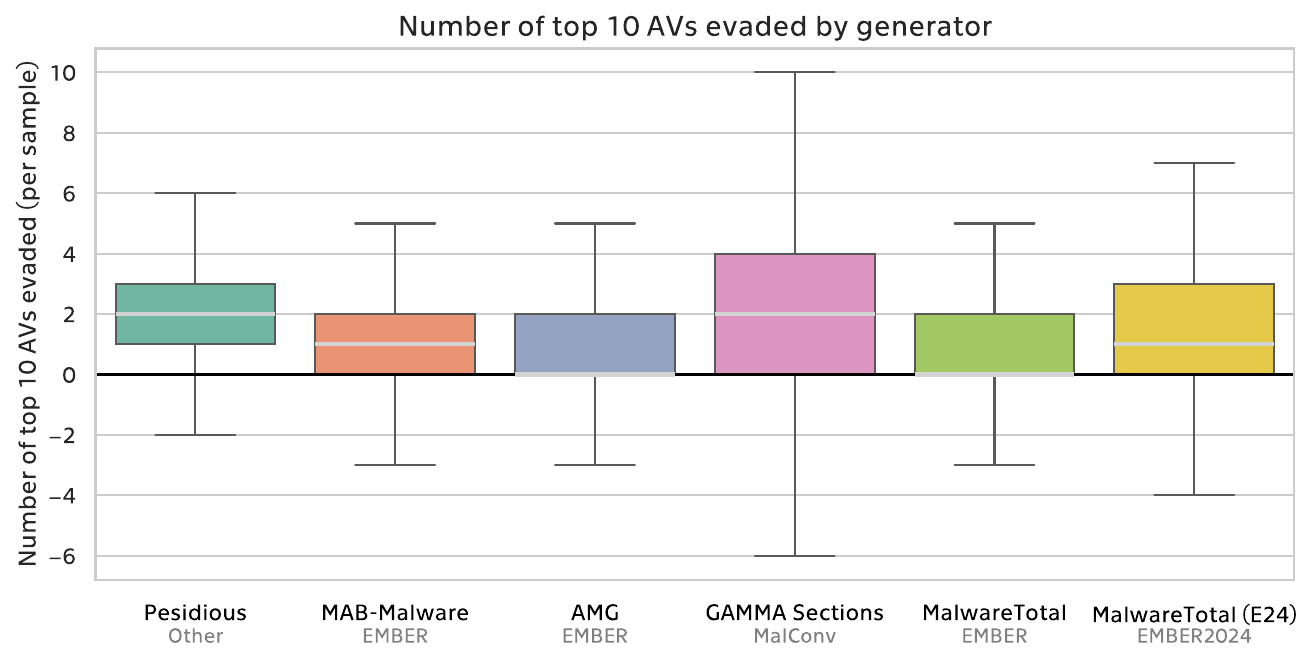}

\caption{Box plot of detection drops by generator for selected top 10 antivirus products on VirusTotal for the family-labelled dataset.}
\label{fig:fttav}
\end{figure}

Most of our top 10 antivirus products exhibited solid resistance to our adversarial attacks. Adversarial malware generators were unable to consistently evade most of them, nevertheless some of the top 10 AVs were less resilient against such attacks. Pesidious had compelling results in this comparison, as did GAMMA Sections. The results of this attack can be seen in Fig.~\ref{fig:fttav}.

\begin{figure}[h]
\includegraphics[width=\textwidth]{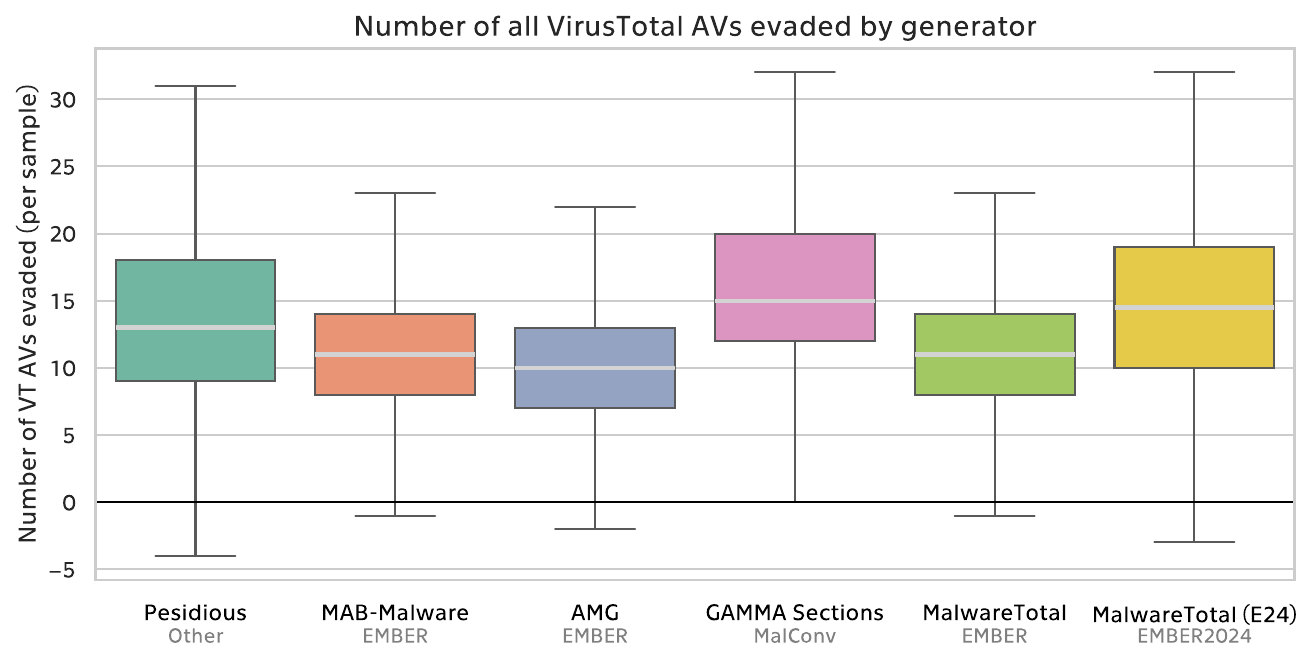}

\caption{Box plot of detection drops across all antivirus products available on VirusTotal for the family-labelled dataset.}
\label{fig:faav}
\end{figure}

Certain antivirus products deployed on VirusTotal are highly vulnerable to our adversarial attacks and many adversarial samples from most generators were able to evade these classifiers with a fairly high consistency. Generators not targeting EMBER exhibit better performance in this experiment. At the time the experiments were conducted, VirusTotal aggregated results from 70 antivirus engines. The results are visualised by Fig.~\ref{fig:faav}.

\begin{figure}[h]
\includegraphics[width=\textwidth]{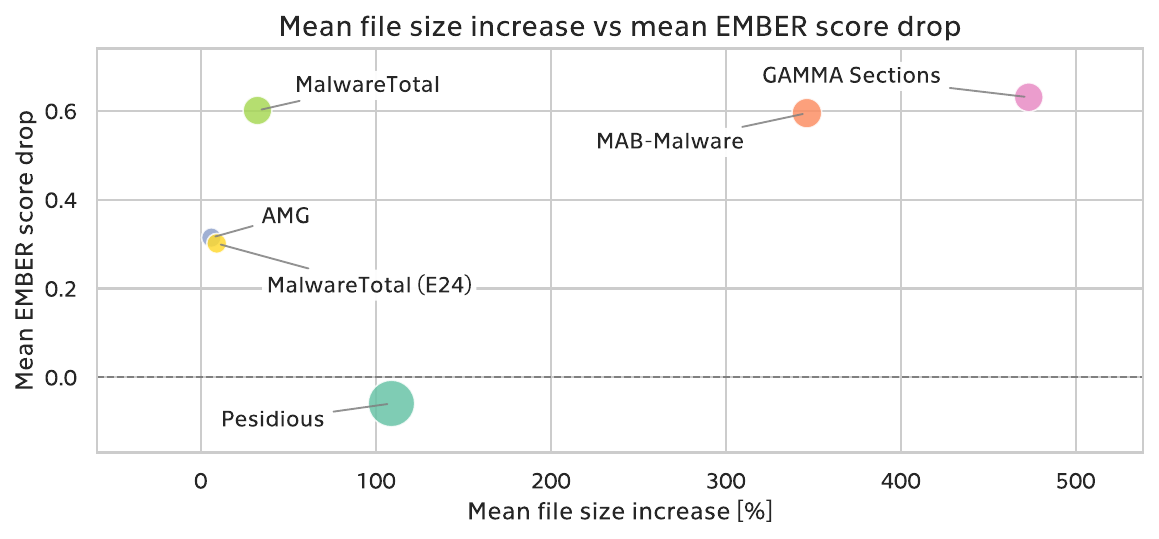}

\caption{Plot of file size increase compared to EMBER score drop for the family-labelled dataset (point size represents number of generated EMBER-evasive samples).}
\label{fig:fse}
\end{figure}

Furthermore we examine how generators differed in terms of output file size increases, as it is an important component of evasion cost. GAMMA Sections is effective in circumventing classification, but tends to generate larger files. MalwareTotal-generated adversarial samples have the best mean ratio of EMBER score drop to file size increase. Figure~\ref{fig:fse} provides an overview for this examination.

\subsubsection{Final dataset statistics}

The source family-labelled dataset contains 44,607 malware samples, 18,750 (42.1\%) of which are classified as malicious by EMBER. The final adversarial family-labelled dataset consists of 44,347 samples, with 18,441 out of the 18,750 originally malicious samples (98.4\%) successfully evading EMBER classification. In total, 41,037 (92.5\%) of the final 44,347 adversarial samples are undetected by EMBER, compared to 25,813 (57.9\%) in the RawMal-TF source dataset. It should be noted that we also used adversarial generators that were not targeting EMBER. As a result, the final dataset contains substantially more samples than the 18,750 originally detected by EMBER, since these additional generators produced modified samples that reduced EMBER detections for many previously undetected files as well. Samples we consider benign have an EMBER score below the threshold of 0.871, which was selected to achieve a FPR below 0.1\%.

However, 309 originally-malicious samples failed to evade detection and remain classified as malicious, representing a 1.7\% failure rate. Additionally, 3,001 originally-benign samples were modified such that their adversarial variants trigger EMBER's detector, a pathological outcome. All but one of these pathological samples were generated by Pesidious, with the remaining one produced by MalwareTotal (E24). Neither of these generators uses EMBER as its target classifier. We still include these pathological samples in the final dataset. Table~\ref{tab:fsbg} provides an overview of generator distribution in the final dataset.

\begin{table}[h]
\caption{Count of samples by generator in the final family-labelled dataset.}
\label{tab:fsbg}
\centering
\begin{tabular*}{\textwidth}{
@{\extracolsep{\fill}}
l
S[table-format=5.0,group-separator={,},group-minimum-digits=4]
S[table-format=2.2]
@{}
}
\hline\noalign{\smallskip}
\textbf{Generator} &
\multicolumn{1}{c}{\textbf{Count}} &
\multicolumn{1}{c}{\textbf{Share (\%)}} \\
\noalign{\smallskip}\svhline\noalign{\smallskip}
Pesidious          & 21069 & 47.51 \\
MalwareTotal       &  9148 & 20.63 \\
GAMMA Sections     &  6434 & 14.51 \\
MalwareTotal (E24) &  3570 &  8.05 \\
MAB-Malware        &  3232 &  7.29 \\
AMG                &   894 &  2.02 \\
\noalign{\smallskip}\hline
\end{tabular*}
\end{table}

\begin{figure}[h]
\includegraphics[width=\textwidth]{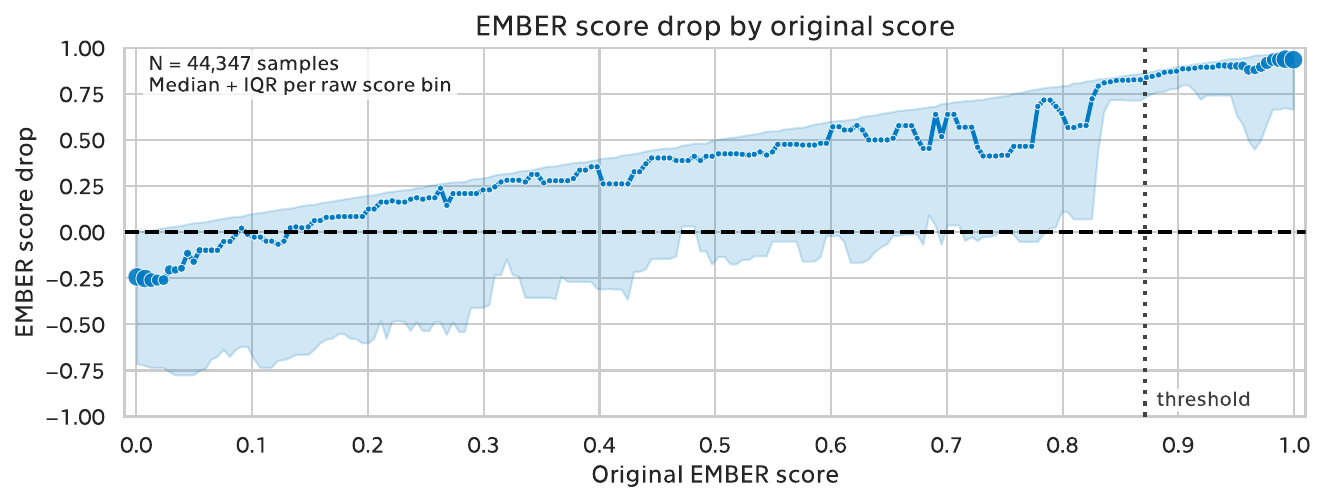}

\caption{Median EMBER score drop versus original score for samples in the final family-labelled dataset. Blue points and line show the median drop per original score bin, with point size proportional to the number of samples in the bin. The shaded blue band shows the interquartile (IQR) range (25th to 75th percentile). Note that the bins on the extreme edges of the score distribution contain the majority of samples.}
\label{fig:f5}
\end{figure}

As evident from Fig.~\ref{fig:f5}, generators were able to significantly decrease the score for the majority of samples. However, due to EMBER's limited recall on our source dataset, there is also a sizeable number of samples with an increased EMBER score. We still consider them valuable; they are provided as mutated samples which might aid further research.

\begin{figure}[h]
\includegraphics[width=\textwidth]{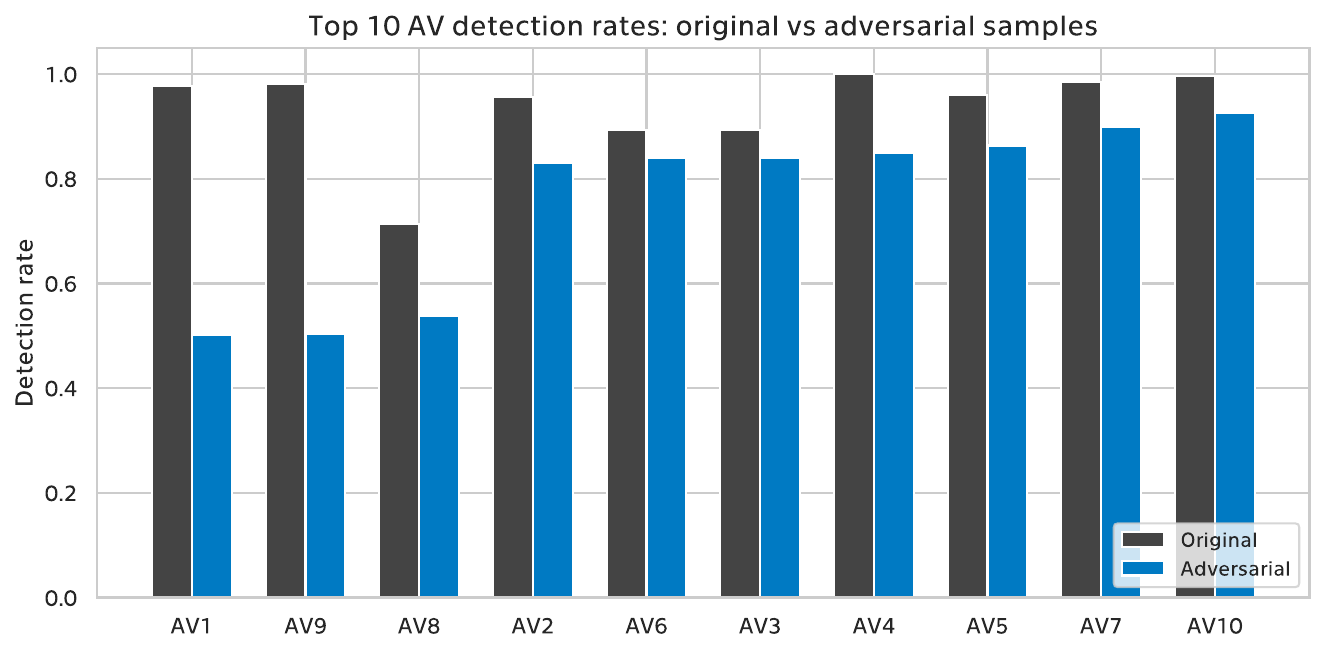}

\caption{Comparison of detection rates for original and adversarial samples by AV vendor in the final family-labelled dataset.}
\label{fig:f6}
\end{figure}

Figure~\ref{fig:f6} highlights that some antivirus engines appear to be fairly vulnerable to adversarial attacks. However, as mentioned before, AV vendors may use different configurations for their software running on VirusTotal. Most of the leading AV products are fairly successful at resisting our attacks.

\subsection{Analysing the type-labelled dataset}

\begin{figure}[h]
\includegraphics[width=\textwidth]{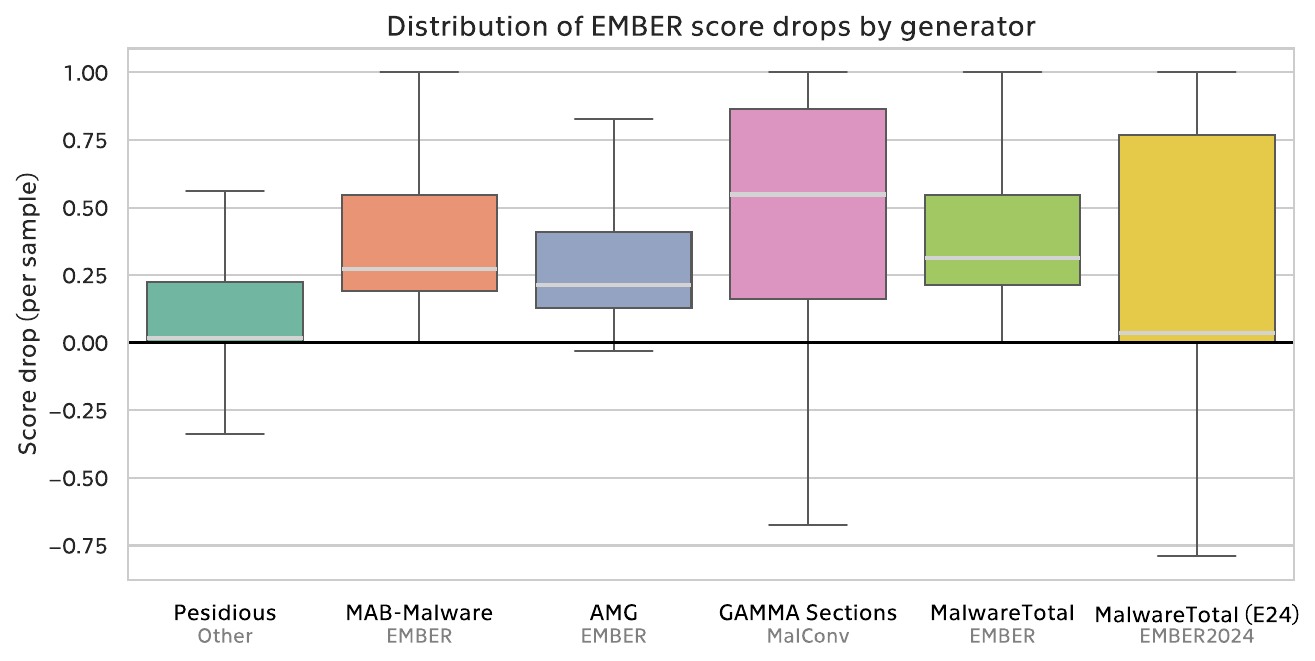}

\caption{Box plot of EMBER score drops by generator for the type-labelled dataset. Note that generators using EMBER as their target never make the score worse, while others might. AMG's result here is slightly anomalous, most likely explained by its oversized samples causing bad classifications.}
\label{fig:temd}
\end{figure}

As with the family-labelled dataset, classifiers trained for EMBER are more effective at evading EMBER, compared to Pesidious and MalwareTotal (E24). However, GAMMA Sections performs notably well here, it achieves the highest EMBER score drop out of all generators. Pesidious generates significantly fewer pathological samples with increased EMBER scores, compared to the family-labelled dataset. These results can be seen in Fig.~\ref{fig:temd}.

\begin{figure}[h]
\includegraphics[width=\textwidth]{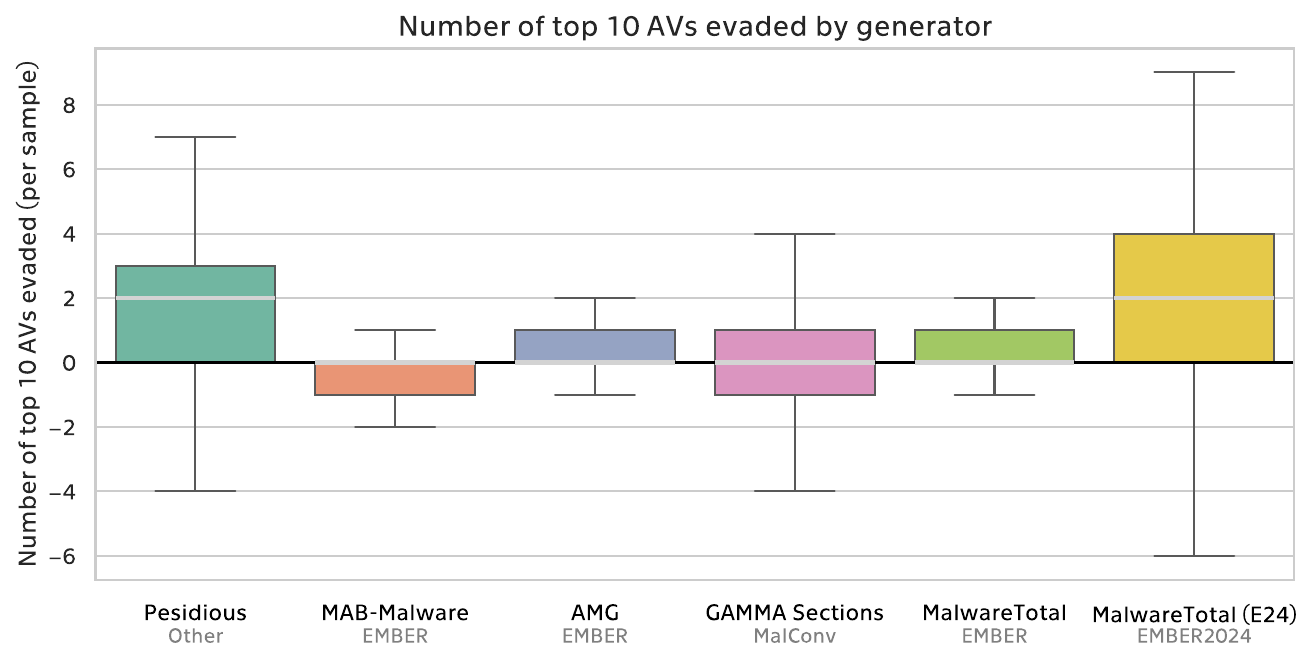}

\caption{Box plot of detection drops by generator for selected top 10 antivirus products on VirusTotal for the type-labelled dataset.}
\label{fig:tttav}
\end{figure}

Most generators for the type-labelled dataset find it harder to evade top antivirus products, compared to the family-labelled dataset. Only MalwareTotal (E24) exhibits better results compared in this statistic. Most generators could not consistently evade top AVs for the type-labelled dataset. This is showcased by Fig.~\ref{fig:tttav}.

\begin{figure}[h]
\includegraphics[width=\textwidth]{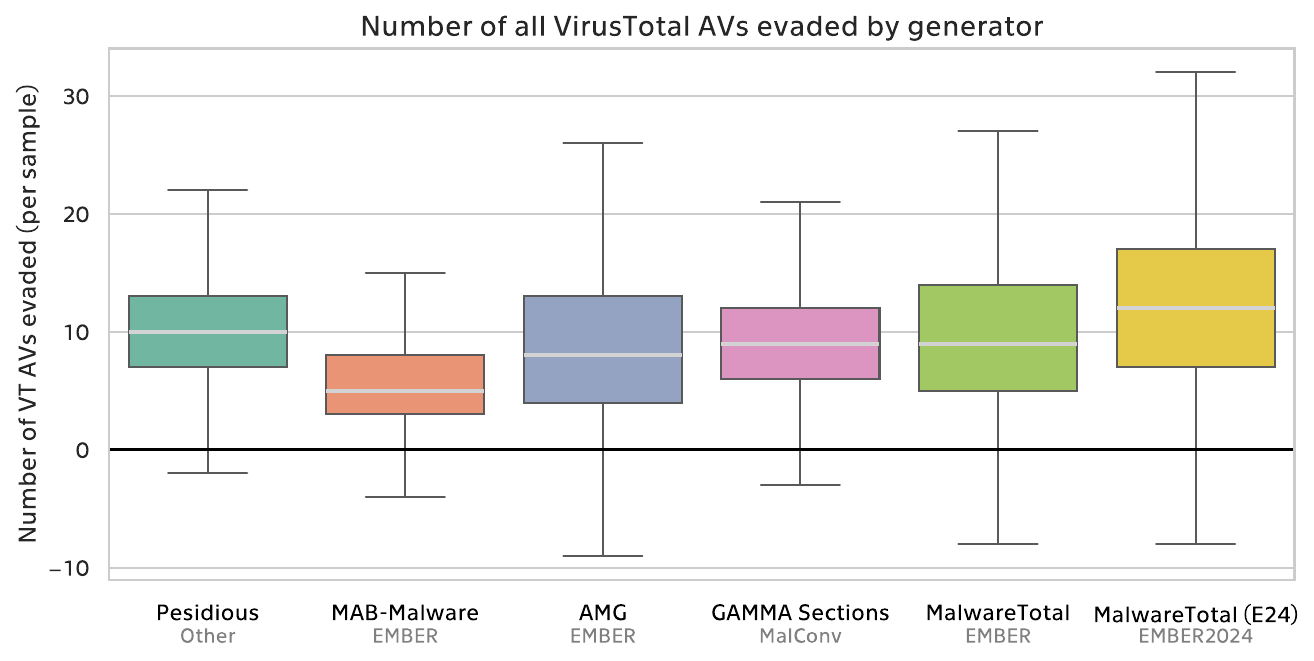}

\caption{Box plot of detection drops across all antivirus products available on VirusTotal for the type-labelled dataset.}
\label{fig:taav}
\end{figure}

Similarly to the previous statistic, Fig.~\ref{fig:taav} showcases that generators were less successful at evading commercial antivirus software in the type-labelled dataset. All of the generators we utilised have a lower score here, compared to the family-labelled dataset.

\begin{figure}[h]
\includegraphics[width=\textwidth]{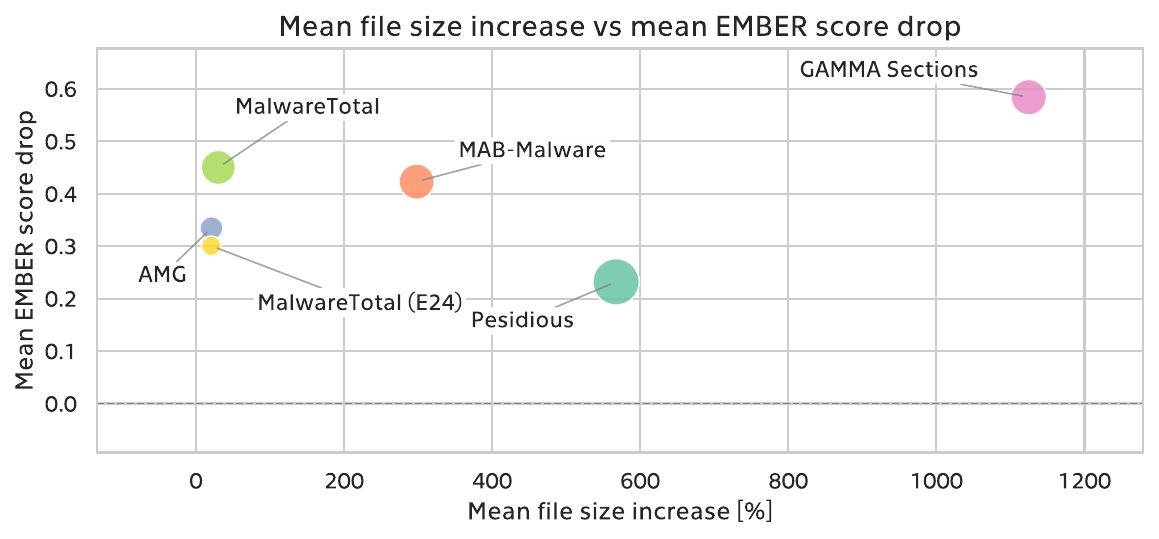}

\caption{Plot of file size increase compared to EMBER score drop for the type-labelled dataset (point size represents number of generated EMBER-evasive samples).}
\label{fig:tes}
\end{figure}

Examining the relationship between file size growth and EMBER scores (see Fig.~\ref{fig:tes}), GAMMA Sections is even more of an outlier compared to the family-labelled dataset, showing a mean file size increase of more than a 1000\%. MAB-Malware's mean size increase drops, while Pesidious' grows. Mean file size increases for AMG, MalwareTotal and MalwareTotal (E24) stay low.

\subsubsection{Final dataset statistics}

The source RawMal-TF type-labelled dataset contains 152,494 files, 40,919 of which are PE files successfully classified by EMBER. Out of valid PE files, 34,327 samples are unique (some samples appear in multiple folders) and serve as our base dataset. 25,204 (73.4\%) of the unique source files are classified as malicious.

The final type-labelled adversarial dataset consists of 33,596 files. Out of the 25,204 EMBER-detected malicious files, 23,239 (92.2\%) were able to be modified to evade the detection. 1,965 samples failed to be modified, indicating a higher failure rate of 7.8\% compared to the family-labelled dataset.

In total, the final type-labelled dataset contains 31,413 (93.5\%) undetected files, while the original unique dataset contains 9,010 (26.2\%). Only 218 samples were turned from benign to malicious. Pesidious generated 205 of these pathological cases, while GAMMA Sections generated 13 such samples. As with the family-labelled dataset, samples considered benign have an EMBER score below the threshold of 0.871. Table~\ref{tab:tcbg} breaks down the final generator distribution.

\begin{table}[h]
\caption{Count of samples by generator in the final type-labelled dataset.}
\label{tab:tcbg}
\centering
\begin{tabular*}{\textwidth}{
@{\extracolsep{\fill}}
l
S[table-format=5.0,group-separator={,},group-minimum-digits=4]
S[table-format=2.2]
@{}
}
\hline\noalign{\smallskip}
\textbf{Generator} &
\multicolumn{1}{c}{\textbf{Count}} &
\multicolumn{1}{c}{\textbf{Share (\%)}} \\
\noalign{\smallskip}\svhline\noalign{\smallskip}
MalwareTotal       & 9943 & 29.60 \\
GAMMA Sections     & 8748 & 26.04 \\
Pesidious          & 8086 & 24.07 \\
MAB-Malware        & 3574 & 10.64 \\
AMG                & 1653 &  4.92 \\
MalwareTotal (E24) & 1592 &  4.74 \\
\noalign{\smallskip}\hline
\end{tabular*}
\end{table}

\begin{figure}[t]
\includegraphics[width=\textwidth]{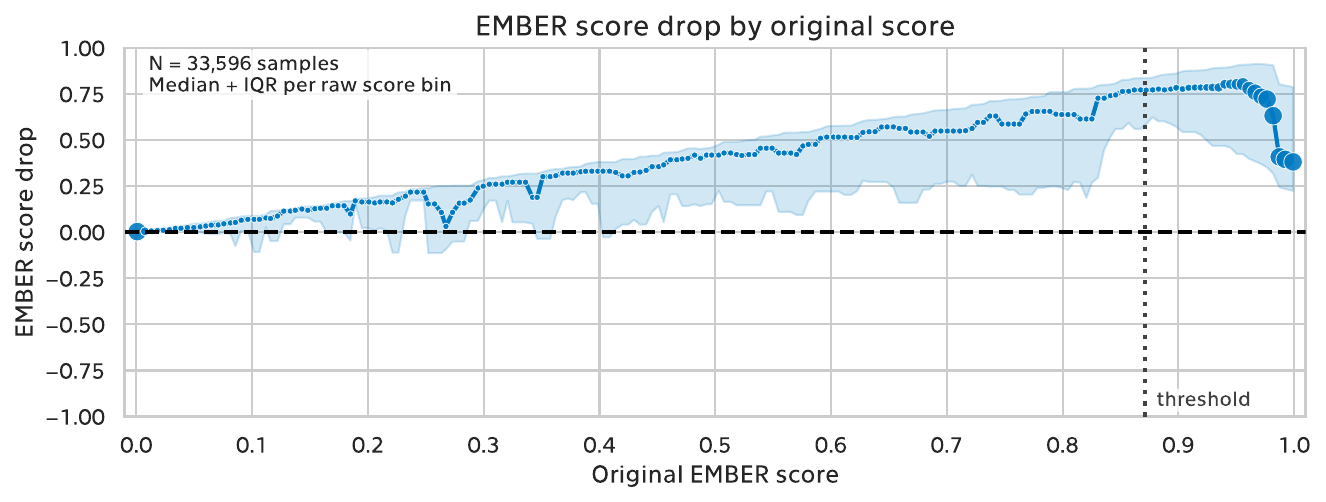}

\caption{Median EMBER score drop versus original score for samples in the final type-labelled dataset. Blue points and line show the median drop per original score bin, with point size proportional to the number of samples in the bin. The shaded blue band shows the interquartile (IQR) range (25th to 75th percentile). Note that the bins on the extreme edges of the score distribution contain the majority of samples.}
\label{fig:tesd}
\end{figure}

Compared to the distribution shown in Fig.~\ref{fig:f5}, the lower end of the score range in Fig.~\ref{fig:tesd} exhibits a noticeably different pattern. We believe this difference is partially explained by EMBER achieving substantially higher recall on the type-labelled dataset than on the family-labelled dataset, resulting in significantly fewer samples exhibiting score increases. Furthermore, we observe a more extreme trend where samples with the highest scores (i.e., highest classifier confidence) have lower score drops. For such samples, it is more difficult for generators to evade EMBER. Although the same effect is also observable in the family-labelled dataset, it is considerably less pronounced there.

\begin{figure}[h]
\includegraphics[width=\textwidth]{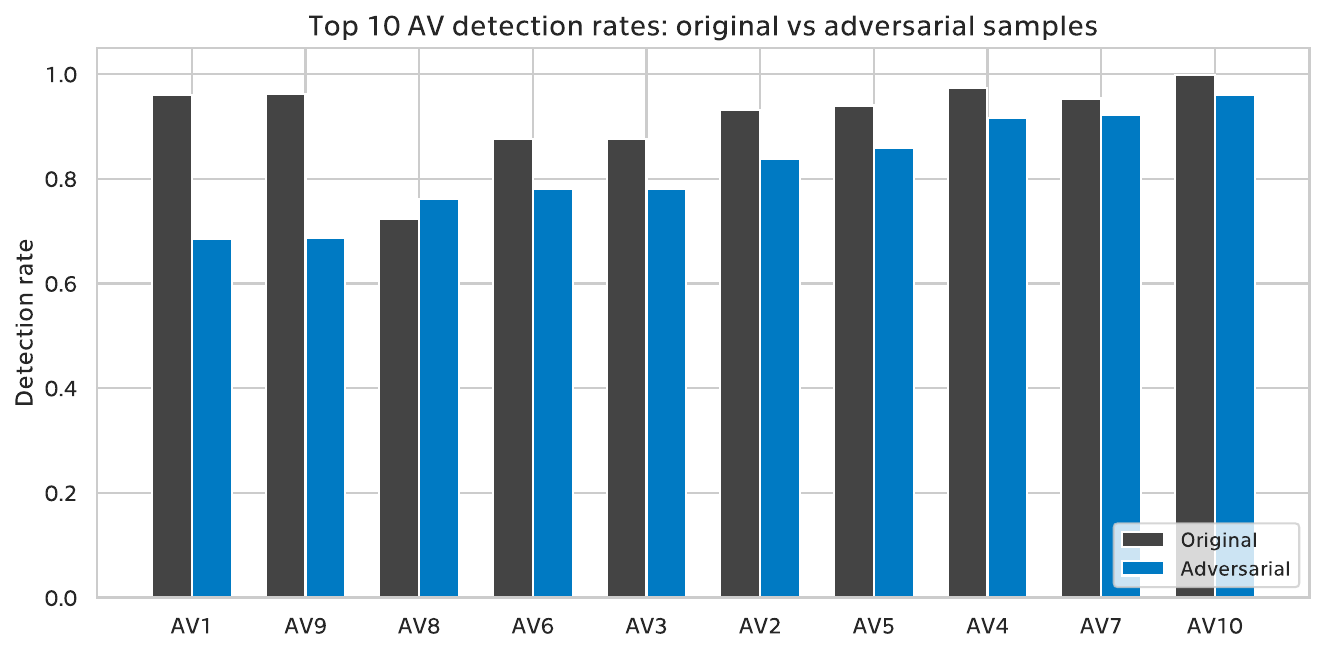}

\caption{Comparison of detection rates for original and adversarial samples by AV vendor in the final type-labelled dataset.}
\label{fig:tttavb}
\end{figure}

In Fig.~\ref{fig:tttavb}, we observe that top 10 AVs detection rate drops are significantly lower compared to the family-labelled dataset. However, this appears to be a side effect of how the resulting samples are picked. Additional per-generator AV detection drop charts show that some generators on their own have lowered detection rates for certain AVs significantly.

\section{Poisoning evaluation}\label{sec:eval}

The final part of this work concerns evaluating the usage of adversarial samples in the classifier training process, and how this affects classifier performance. Our training pipeline is based on EMBER, we are using the same feature set and the same classifier architecture (with a LightGBM-based gradient boosted model). However, we are using a different training dataset (EMBER features provided by RawMal-TF), which is then poisoned (in several different configurations) with features extracted from our adversarial samples.

\subsection{Setup}

This experiment simulates adversarial samples being released into the wild, and classifiers being trained on such samples with incorrect labels. Poisoning evaluation is set up as follows: $C_T$ is the group of original target classifiers, i.e., the ones that adversarial generators are targeting. Let $C_D$ be a classifier not in $C_T$ (i.e., $C_D \notin C_T$). We then train an updated classifier $C_D'$ with training data labelled by the output of $C_D$. Let $\tau = T(C_T, C_D)$, $0 \le \tau \le 1$ be the transferability rate (i.e., the proportion of samples which successfully evaded $C_T$, that also evade $C_D$).

This setup models a scenario where an independent classifier is used to generate labels for adversarial samples being incorporated into a training dataset. We simulate a $\tau$-transferability case by mislabelling $\tau$-fraction of injected malicious samples as benign. In other words, the transferability rate is the fraction of mislabelled samples added to training data: $\tau=0$ means all adversarial samples are correctly labelled as malicious (i.e., in essence adversarial training), while with $\tau=1$ all adversarial samples are incorrectly labelled as benign (i.e., maximum possible poisoning).

We will perform our experiments with various percentages of adversarial samples added to training data, as well as differing values of the transferability rate $\tau$. When adding adversarial samples to the training set, we will also ensure class balance by removing samples from the existing dataset with opposite labels.

\subsection{Methodology}

Before training, we adjusted the range of LightGBM hyperparameters and performed a grid search to optimise for the best values (see Table~\ref{tab:hpgrid}). Our dataset is significantly smaller, therefore we greatly reduced the number of tree leaves. Unlike the original EMBER configuration, we chose Stratified K-Fold instead of Time series split, due to missing data about when samples were first discovered. This may lead to inflated benchmark numbers compared to classifiers' real-world performance. However, this does not affect the validity of our conclusions, as our experiments rely on relative comparisons rather than absolute benchmarks.

\begin{table}[h]
\caption{Hyperparameter grids used for LightGBM classifier tuning. Dashes (---) indicate parameters not present in that configuration.}
\label{tab:hpgrid}
\centering
\begin{tabular*}{\textwidth}{@{\extracolsep{\fill}}lll@{}}
\hline\noalign{\smallskip}
\textbf{Parameter} &
\textbf{Final grid} &
\textbf{Original grid} \\
\noalign{\smallskip}\svhline\noalign{\smallskip}
Iterations & --- & 500, 1000 \\
Learning rate & 0.02, 0.05 & 0.005, 0.05 \\
Number of leaves & 31, 63, 127 & 256, 512, 1024 \\
Max depth & $-1$, 8 & --- \\
Min data in leaf & 50, 100, 200 & --- \\
Feature fraction & 0.6, 0.8 & 0.5, 0.8, 1.0 \\
Bagging fraction & 0.7, 0.9 & 0.5, 0.8, 1.0 \\
Bagging frequency & 5 & --- \\
L2 regularisation ($\lambda$) & 0.0, 1.0 & --- \\
\noalign{\smallskip}\hline
\end{tabular*}
\end{table}

A baseline model serves as a reference point for comparing different configurations. We train the baseline model on EMBER feature vectors from RawMal-TF with the provided training/testing split without modifications except for data cleaning and merging. For poisoning configurations, we modify the training dataset while leaving the testing dataset intact. This testing dataset is used to calculate the F1 score, recall, precision and accuracy.

Features for adversarial samples are extracted using the EMBER pipeline from the final adversarial dataset binaries for both datasets respectively. These feature vectors are then split into a training and testing set, where the training set serves as a pool from which adversarial samples are added to the original training data. The testing set is used for evaluating the evasion rate (i.e., the share of adversarial samples evading the poisoned classifier).

Note that this poisoning evaluation experiment was performed before the faulty Pesidious samples were discovered (see Sect.~\ref{sec:metho}). While these samples are removed from the final dataset, they may have affected the poisoning results slightly. However, we believe the impact on results to be minor, as these samples manifested primarily as duplicate feature vectors in the training data, which are unlikely to have significantly skewed the results.

\subsection{Results}\label{sec:pr}
For each dataset, we trained a total of 99 classifiers ($11 \times 9$), covering all combinations of mislabelled sample rates (transferability rate $\tau$) and poisoned fractions listed in Table~\ref{tab:pconf}. The performance of each trained classifier was then systematically evaluated. We report results using the F1 score, which is defined as follows:

\begin{equation}
F_1 = \frac{2 \times \text{precision} \times \text{recall}}{\text{precision} + \text{recall}}
\end{equation}

\begin{table}[h]
\caption{Configurations used for poisoning experiments.}
\label{tab:pconf}
\centering
\begin{tabular*}{\textwidth}{@{\extracolsep{\fill}}ll@{}}
\hline\noalign{\smallskip}
\textbf{Mislabelled samples ($\tau$)} &
0.0, 0.1, 0.2, 0.3, 0.4, 0.5, 0.6, 0.7, 0.8, 0.9, 1.0 \\
\textbf{Poisoned fraction} &
0.001, 0.0025, 0.005, 0.01, 0.02, 0.03, 0.05, 0.1, 0.2 \\
\noalign{\smallskip}\hline
\end{tabular*}
\end{table}
In Figs.~\ref{fig:fpeheat}--\ref{fig:tpfheat}, we report all metrics as fractions in $[0,1]$. In text, classification metrics (F1, precision, recall) remain in $[0,1]$, while evasion rates are reported as percentages for readability.

\subsubsection{Family-labelled dataset}\label{sec:fldp}

The baseline classifier variant (i.e., with poisoned fraction $= 0$) trained on the family-labelled dataset achieves an evasion rate of 26.13\%, where the evasion rate denotes the percentage of adversarial malicious samples classified as benign by the baseline model. At the same time, the classifier attains an F1 score of 0.9960, with recall and precision values of 0.9922 and 0.9999, respectively. These results appear strong at first glance, however we believe they are best explained by this dataset's low diversity. The family-labelled dataset contains a relatively small number of malware families with a significant amount of similar samples, leading to overfitting in the classifier. Nevertheless, this experiment still produces valuable data.

\begin{figure}[t]
\includegraphics[width=0.95\textwidth]{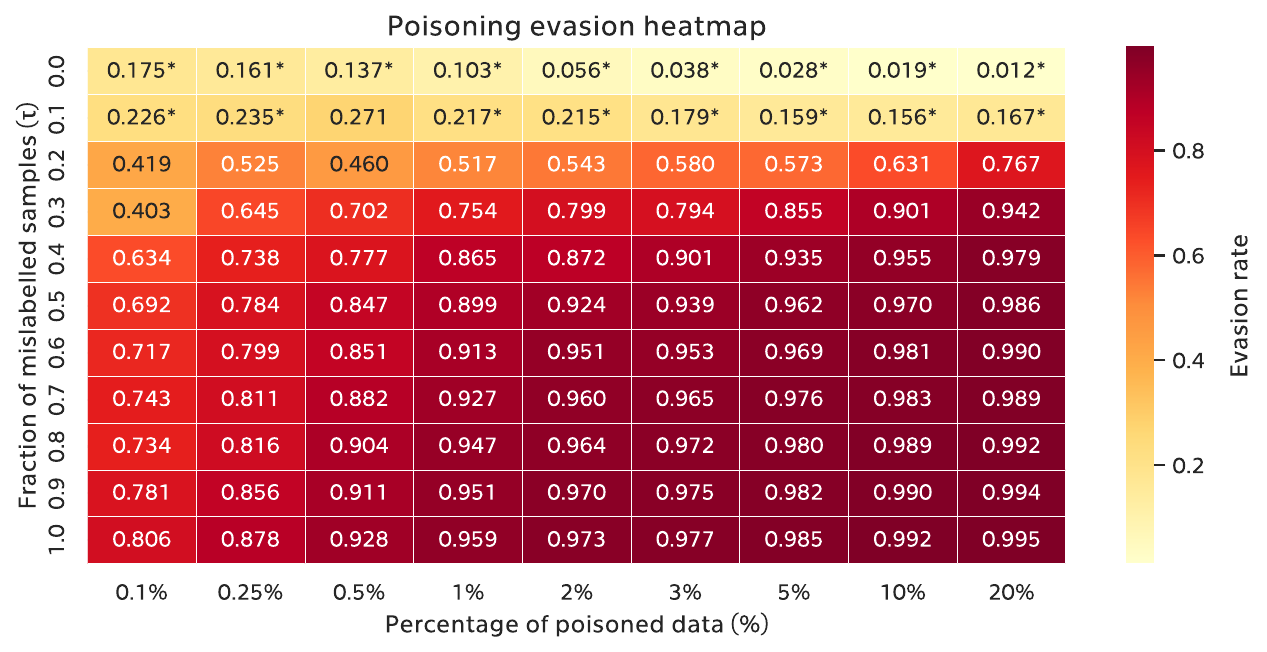}

\caption{Family-labelled dataset: heatmap of evasion rates under poisoning attacks. Note: values with evasion rates lower than the baseline are marked with an asterisk (*).}
\label{fig:fpeheat}
\end{figure}

\begin{figure}[t]
\includegraphics[width=0.95\textwidth]{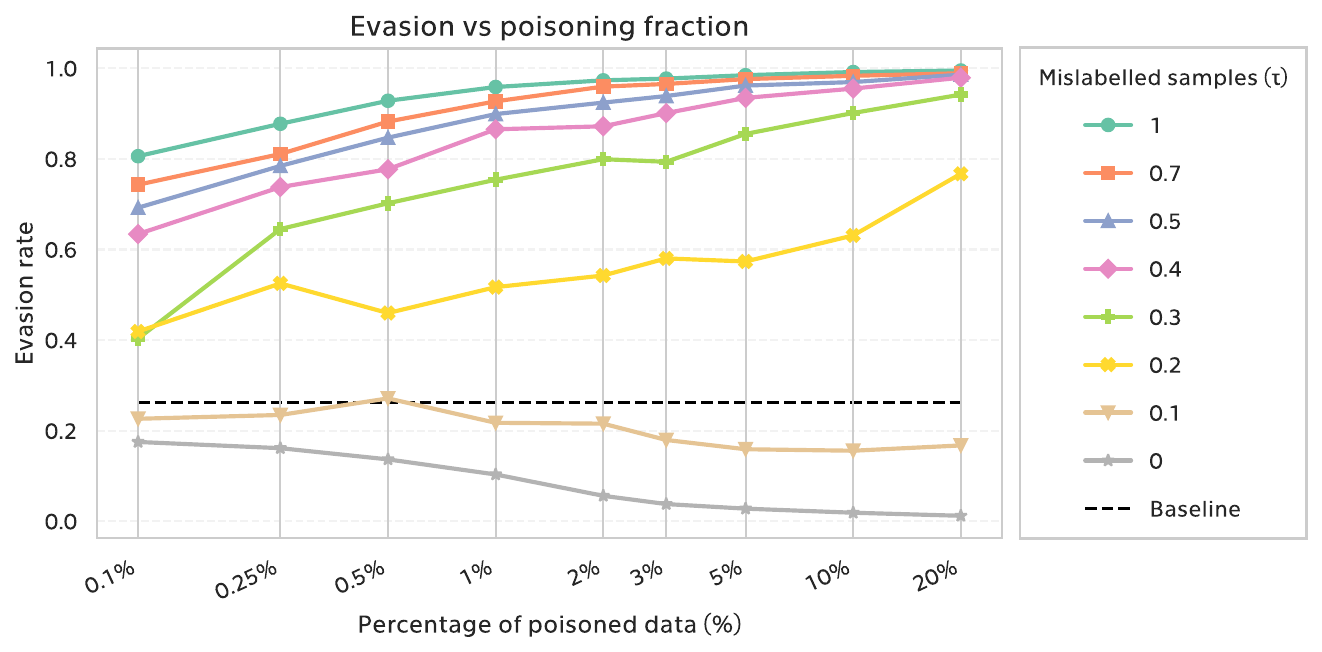}

\caption{Family-labelled dataset: evasion rate by poisoning fraction and threshold $\tau$. Note: some values of $\tau$ are omitted for better readability.}
\label{fig:fpeline}
\end{figure}

In Fig.~\ref{fig:fpeheat} and Fig.~\ref{fig:fpeline}, we observe that $\tau=0$ and $\tau=0.1$ in fact improve the classifier's ability to correctly detect adversarial samples as malicious, with any fraction of poisoned data in the training dataset (except for minor fluctuations).
For any $\tau\ge0.2$, we note that the evasion rate is significantly higher, again for any tested fraction of poisoned data. With increasing values of $\tau$ and increasing fractions of poisoned data, the evasion rate increases fairly consistently.

\begin{figure}[h]
\includegraphics[width=\textwidth]{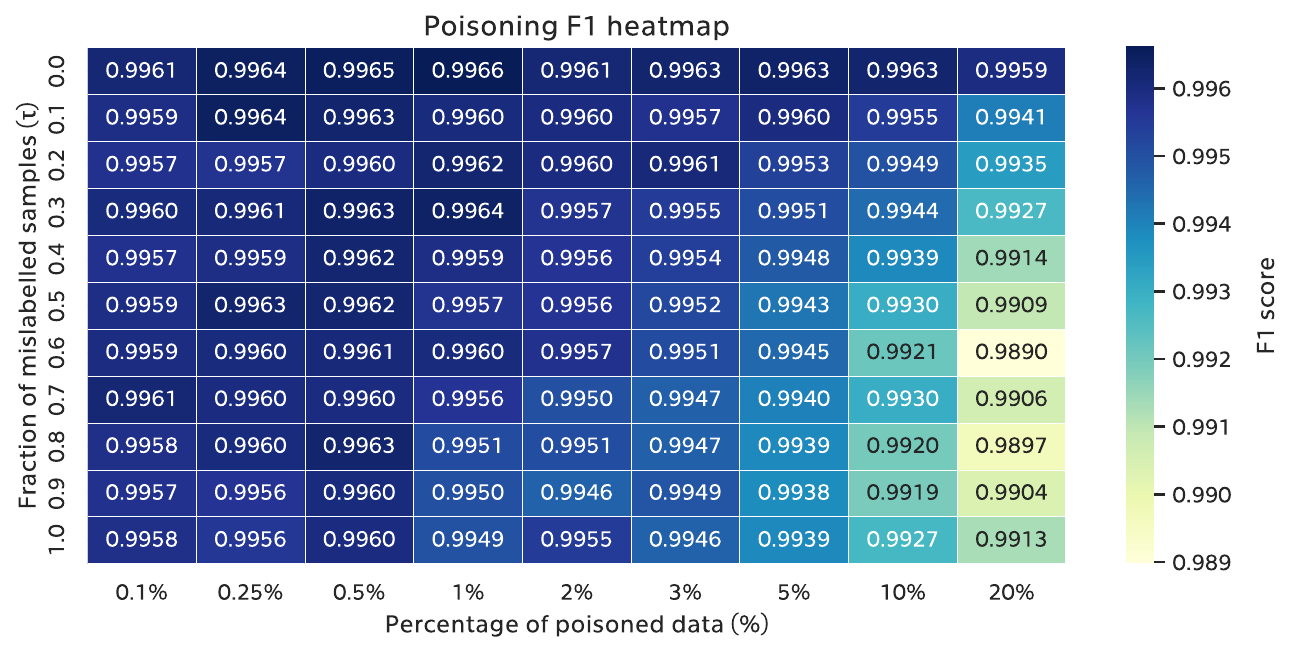}

\caption{Family-labelled dataset: heatmap of F1 scores under poisoning attacks.}
\label{fig:fpfheat}
\end{figure}

Figure~\ref{fig:fpfheat} clearly demonstrates that F1 scores for poisoned classifiers remain remarkably consistent across most conditions, only showing noticeable drops at high values of~$\tau$ combined with high fractions of poisoned data. The limited impact of poisoning on overall classification performance is a particularly concerning finding: a poisoned classifier continues to perform well on clean data, making it difficult to detect that the training dataset contains mislabelled samples without targeted evaluation specifically designed to surface this issue. This highlights the importance of careful dataset construction and suggests that careful inclusion of correctly labelled adversarial samples helps in significantly lowering evasion rates, while mislabelling such samples can have a devastating effect on classifier performance.

\subsubsection{Type-labelled dataset}

Compared to the family-labelled classifier, the type-labelled dataset's baseline classifier achieves scores more in line with real-world expectations, with an F1 score of 0.9871 (recall 0.9746, precision 0.9998). Its evasion rate is also markedly higher at 72.9\%. We believe this is explained by this dataset's greater diversity.

\begin{figure}[t]
\includegraphics[width=0.95\textwidth]{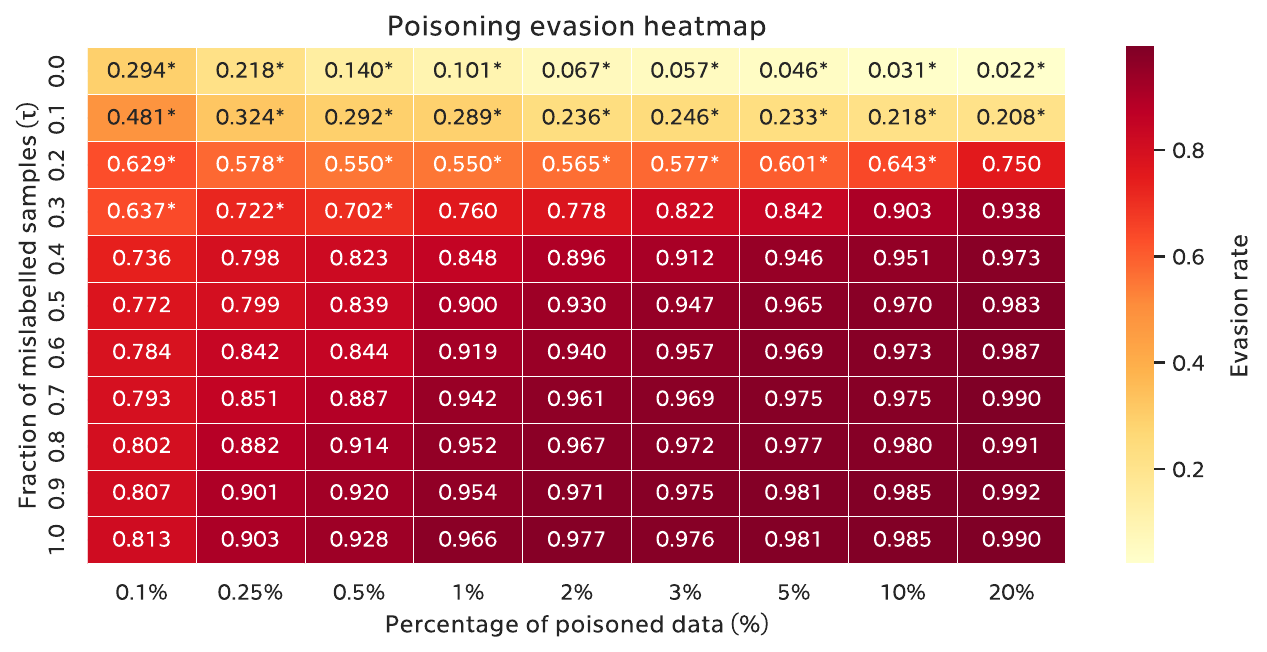}

\caption{Type-labelled dataset: heatmap of evasion rates under poisoning attacks. Note: values with evasion rates lower than the baseline are marked with an asterisk (*).}
\label{fig:tpeheat}
\end{figure}

\begin{figure}[t]
\includegraphics[width=0.95\textwidth]{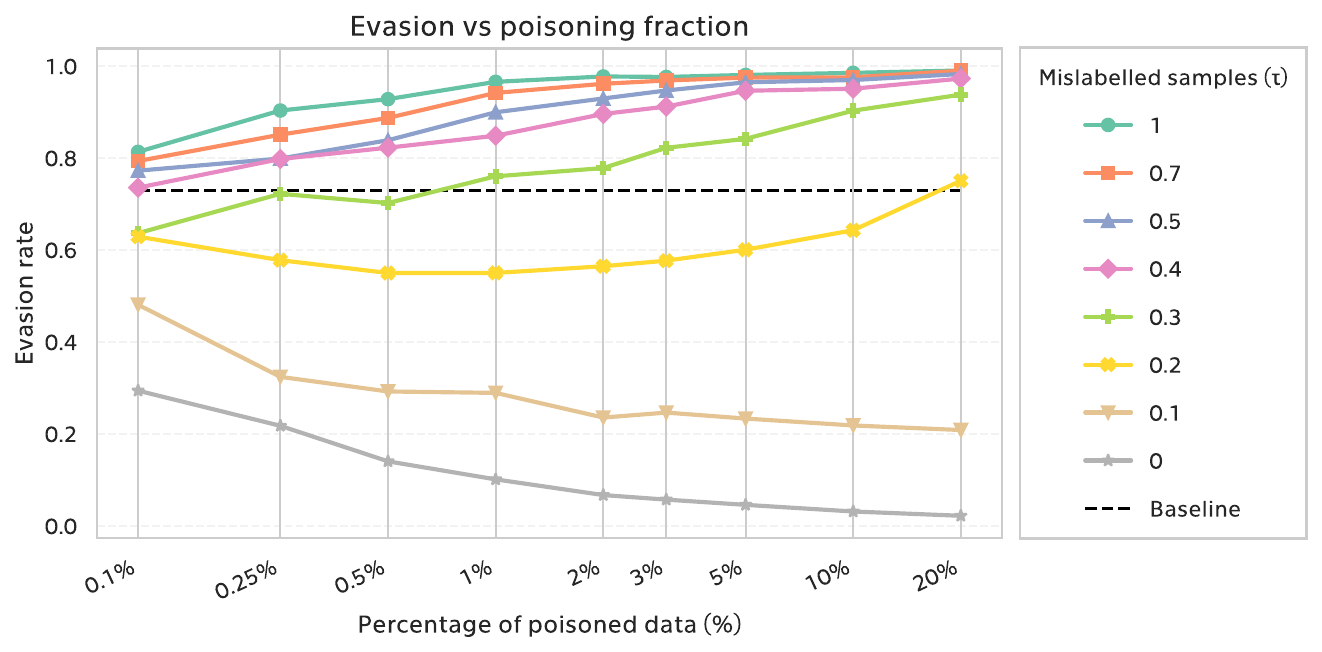}

\caption{Type-labelled dataset: evasion rate by poisoning fraction and threshold $\tau$. Note: some values of $\tau$ are omitted for better readability.}
\label{fig:tpeline}
\end{figure}

Figure~\ref{fig:tpeheat} and Fig.~\ref{fig:tpeline} indicate that values of $\tau=0$, $\tau=0.1$, $\tau=0.2$ and in part $\tau=0.3$ lead to an improvement in the classifier's ability to detect adversarial samples.

As shown in Fig.~\ref{fig:tpeline}, for lower $\tau$ values this effect increases with growing fractions of poisoned data, while for higher values a threshold is crossed and poisoning starts to negatively impact the classifier's performance. While the baseline evasion rate is relatively high, it increases even more with higher $\tau$ values, even at low fractions of poisoned data.

\begin{figure}[h]
\includegraphics[width=\textwidth]{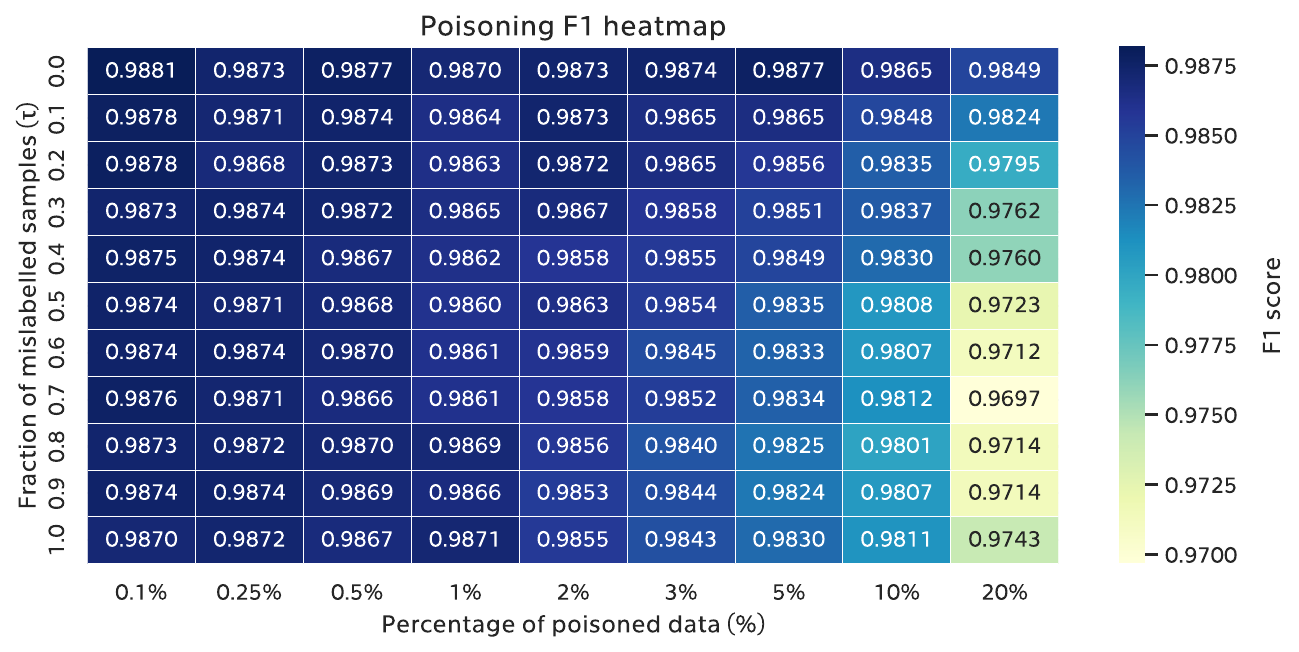}

\caption{Type-labelled dataset: heatmap of F1 scores under poisoning attacks.}
\label{fig:tpfheat}
\end{figure}

Figure~\ref{fig:tpfheat} showcases that F1 scores still stay relatively stable with lower $\tau$ values and lower fractions of poisoned data, but start to show greater deviation at higher values. As discussed in Sect.~\ref{sec:fldp}, this may lead to classifier degradation, unless special care is taken during classifier training.

\subsection{Cross-evaluation and summary}

We also performed a cross-evaluation experiment (i.e., classifying the family-labelled dataset's test set with the type-labelled dataset's baseline classifier and vice versa) to help explain the anomalous scores for the family-labelled dataset.

Cross-evaluating the type-labelled classifier (on family-labelled data), we get an F1 score of 0.9323 (recall 0.8771, precision 0.9950). This result demonstrates that while this classifier is far from optimal, it performs within reasonable expectations, given the dataset is relatively small and no significant optimisations were undertaken.

However, running a cross-evaluation on the family-labelled classifier yields an F1 score of 0.4317 (recall 0.2753, precision 0.9995), demonstrating a poor result in detecting malicious samples (i.e., its recall is very low). This result hints at problems with training the family-labelled classifier, likely due to a limited diversity of samples.

To summarise the poisoning attack results, despite the limitations of our experimental setup, we demonstrated that poisoning attacks can be highly impactful even when only relatively small fractions of the training data are poisoned and transferability rates remain low. Conversely, poisoning with very low $\tau$ values effectively acted as a form of adversarial training in our experiments and contributed to reduced evasion rates. Overall, this investigation highlights the importance of careful dataset construction and suggests that the deliberate inclusion of correctly labelled adversarial samples can significantly improve classifier robustness against evasion attacks.

\FloatBarrier
\section{Conclusion}\label{sec:conc}

The main goal of this work was to create a well-annotated adversarial malware dataset, suitable for use in further research. The dataset provides researchers with diverse adversarial PE samples and rich metadata for use in adversarial malware research and classifier robustness evaluation. The published dataset, including binaries and accompanying metadata, is available at \url{https://github.com/CS-and-AI/AdvMal-TF}.

The family-labelled adversarial dataset contains 44,347~samples built from 106,394 candidates, while the type-labelled dataset contains 33,596 samples selected from 96,966 unique generated candidates. The family-labelled and the type-labelled datasets feature 98.35\% and 92.20\% EMBER-evasive samples of originally EMBER-detected samples, respectively.

To generate the dataset, we built a container-based pipeline to run multiple adversarial malware generators at scale, in parallel over the server's CPUs. The pipeline's tooling allowed us to monitor generator operation, control resource usage on a multi-user server and isolate failures (e.g., problematic files causing process hangs). Since many original samples produced multiple adversarial variants across generators, we designed a selection algorithm to pick the best variant per original file based on the EMBER classifier score, while taking file size (evasion cost) into account.

To support this selection process and to create the plots in Sect.~\ref{sec:analysis}, we created scripts to classify all adversarial samples with EMBER, EMBER2024 and VirusTotal. The EMBER and EMBER2024 scripts run in parallel to optimise resource usage, while the VirusTotal script automatically saves its intermediate state to disk to accommodate daily VT API limits, enabling classification of a directory over the course of multiple days. We also used the resulting dataset to perform data poisoning experiments.

A recurring challenge throughout this work was getting adversarial malware generators to function reliably. Most generators featured outdated dependencies or were designed for a specific system environment, and some exhibited unexpected reliability issues. To proceed with the dataset creation, we had to mitigate such problems. Furthermore, to enhance the diversity of the dataset, we modified the MalwareTotal generator to target EMBER2024 in addition to EMBER. Evading EMBER2024 proved challenging for all considered generators, and more broadly, the currently available generators had limited success in evading more advanced classifiers. This suggests a need for further research into adversarial evasion techniques.

The final part of this work concerns data poisoning evaluation. We trained 99 classifiers per dataset across combinations of transferability rate $\tau$ and poisoned data fractions from 0.1\% to 20\%, using the EMBER pipeline with a LightGBM model on RawMal-TF data. The results are unambiguous: poisoning is highly effective even at small scales. On the family-labelled dataset, adding just 0.5\% fully mislabelled samples ($\tau=1.0$) drove the evasion rate from a baseline of 26.1\% up to 92.8\%. The type-labelled dataset, with a higher baseline evasion of 72.9\%, showed analogous sensitivity to high $\tau$ values. $F_1$ scores remained deceptively stable throughout, masking the underlying collapse in adversarial detection.

A notable positive finding is that including correctly labelled adversarial samples (low $\tau$) consistently acted as adversarial training, reducing evasion rates below baseline on both datasets. This shows that label quality matters as much as volume: the same adversarial artefacts that devastate a classifier when mislabelled can harden it when labelled correctly. Together, these results show that robust malware classifiers require diverse, well-curated training data.

Our cross-dataset evaluation emphasised the difference between the two dataset variants. The type-labelled classifier maintained reasonable performance (recall 0.877) when applied to family-labelled test data, while the family-labelled classifier severely degraded on type-labelled data (recall 0.275), confirming that its high baseline $F_1$ is partly an artefact of overfitting to a narrow feature distribution.

We acknowledge that many of the samples in our dataset are only evasive towards less robust classifiers. Nevertheless, the dataset includes samples from multiple generators with varying evasion strategies, and the accompanying metadata can be searched and filtered for any desired properties, allowing researchers to select subsets suited to their needs.

Machine learning classifiers are a crucial part of modern malware detection, and adversarial attacks are among the key problems in ML-based malware classification. Adversarial training with evasive samples is one of the ways to improve classifier robustness against such attacks. We hope this work aids defenders with adversarial training, and makes a valuable contribution to the adversarial malware discourse.

\begin{acknowledgement}
We would like to acknowledge VirusTotal for providing academic API access.
\end{acknowledgement}

\bibliographystyle{spmpsci}
\bibliography{works}

\begin{thebibliography}{10}
\providecommand{\url}[1]{{#1}}
\providecommand{\urlprefix}{URL }
\expandafter\ifx\csname urlstyle\endcsname\relax
  \providecommand{\doi}[1]{DOI~\discretionary{}{}{}#1}\else
  \providecommand{\doi}{DOI~\discretionary{}{}{}\begingroup
  \urlstyle{rm}\Url}\fi

\bibitem{gym}
Anderson, H.S., Kharkar, A., Filar, B., Evans, D., Roth, P.: Learning to evade
  static {PE} machine learning malware models via reinforcement learning
  (2018).
\newblock \doi{10.48550/arXiv.1801.08917}

\bibitem{ember}
Anderson, H.S., Roth, P.: {EMBER}: An open dataset for training static {PE}
  malware machine learning models (2018).
\newblock \doi{10.48550/arXiv.1804.04637}

\bibitem{avtest}
{AV-TEST Institute}: Test antivirus software for {Windows} 11 -- {February}
  2026 (2026).
\newblock \urlprefix\url{https://www.av-test.org/en/antivirus/home-windows}

\bibitem{sgn}
Balc{\i}, E.: {SGN}: Shikata ga nai encoder.
\newblock \urlprefix\url{https://github.com/egebalci/sgn}

\bibitem{rawmal}
Bálik, D., Jureček, M., Stamp, M.: {RawMal-TF}: Raw malware dataset labeled
  by type and family (2025).
\newblock \doi{10.48550/arXiv.2506.23909}

\bibitem{secmlm}
Demetrio, L., Biggio, B.: secml-malware: A {Python} library for adversarial
  robustness evaluation of {Windows} malware classifiers (2021).
\newblock \doi{10.48550/arXiv.2104.12848}

\bibitem{gamma}
Demetrio, L., Biggio, B., Lagorio, G., Roli, F., Armando, A.:
  Functionality-preserving black-box optimization of adversarial {Windows}
  malware.
\newblock IEEE Trans. Inf. Forensics Secur. \textbf{16}, 3469--3478 (2021).
\newblock \doi{10.1109/tifs.2021.3082330}

\bibitem{grosse}
Grosse, K., Papernot, N., Manoharan, P., Backes, M., McDaniel, P.: Adversarial
  examples for malware detection.
\newblock In: S.N. Foley, D.~Gollmann, E.~Snekkenes (eds.) Computer Security --
  ESORICS 2017, pp. 62--79. Springer International Publishing, Cham (2017).
\newblock \doi{10.1007/978-3-319-66399-9_4}

\bibitem{mt}
He, S., Fu, C., Hu, H., Chen, J., Lv, J., Jiang, S.: {MalwareTotal}:
  Multi-faceted and sequence-aware bypass tactics against static malware
  detection.
\newblock In: Proceedings of the IEEE/ACM 46th International Conference on
  Software Engineering, ICSE '24. Association for Computing Machinery, New
  York, NY, USA (2024).
\newblock \doi{10.1145/3597503.3639141}

\bibitem{claravy}
Joyce, R.J., Everett, D., Fuchs, M., Raff, E., Holt, J.: {ClarAVy}: A tool for
  scalable and accurate malware family labeling.
\newblock In: Companion Proceedings of the ACM on Web Conference 2025, WWW '25,
  p. 277–286. Association for Computing Machinery, New York, NY, USA (2025).
\newblock \doi{10.1145/3701716.3715212}

\bibitem{ember24}
Joyce, R.J., Miller, G., Roth, P., Zak, R., Zaresky-Williams, E., Anderson, H.,
  Raff, E., Holt, J.: {EMBER2024} - a benchmark dataset for holistic evaluation
  of malware classifiers.
\newblock In: Proceedings of the 31st ACM SIGKDD Conference on Knowledge
  Discovery and Data Mining V.2, KDD '25, p. 5516–5526. Association for
  Computing Machinery, New York, NY, USA (2025).
\newblock \doi{10.1145/3711896.3737431}

\bibitem{amg}
Kozák, M., Jureček, M., Stamp, M., Troia, F.D.: Creating valid adversarial
  examples of malware.
\newblock J. Comput. Virol. Hacking Tech. \textbf{20}(4), 607--621 (2024).
\newblock \doi{10.1007/s11416-024-00516-2}

\bibitem{comp}
Louthánová, P., Kozák, M., Jureček, M., Stamp, M., Di~Troia, F.: A
  comparison of adversarial malware generators.
\newblock J. Comput. Virol. Hacking Tech. \textbf{20}(4), 623--639 (2024).
\newblock \doi{10.1007/s11416-024-00519-z}

\bibitem{moser}
Moser, A., Kruegel, C., Kirda, E.: Limits of static analysis for malware
  detection.
\newblock In: Twenty-Third Annual Computer Security Applications Conference
  (ACSAC 2007), pp. 421--430 (2007).
\newblock \doi{10.1109/ACSAC.2007.21}

\bibitem{malconv}
Raff, E., Barker, J., Sylvester, J., Brandon, R., Catanzaro, B., Nicholas, C.:
  Malware detection by eating a whole {EXE} (2017).
\newblock \doi{10.48550/arXiv.1710.09435}

\bibitem{ppoa}
Schulman, J., Wolski, F., Dhariwal, P., Radford, A., Klimov, O.: Proximal
  policy optimization algorithms (2017).
\newblock \doi{10.48550/arXiv.1707.06347}

\bibitem{schultz}
Schultz, M.G., Eskin, E., Zadok, E., Stolfo, S.J.: {Data Mining Methods for
  Detection of New Malicious Executables}.
\newblock In: Proceedings 2001 IEEE Symposium on Security and Privacy, p. 0038.
  IEEE Computer Society, Los Alamitos, CA, USA (2001).
\newblock \doi{10.1109/SECPRI.2001.924286}

\bibitem{pezor}
Soncina, F.: {PEzor} (2020).
\newblock
  \urlprefix\url{https://web.archive.org/web/20240314053756/https://iwantmore.pizza/posts/PEzor.html}

\bibitem{pezorgh}
Soncina, F.: {PEzor} (2023).
\newblock \urlprefix\url{https://github.com/phra/PEzor}

\bibitem{mabm}
Song, W., Li, X., Afroz, S., Garg, D., Kuznetsov, D., Yin, H.: {MAB-Malware}: A
  reinforcement learning framework for attacking static malware classifiers.
\newblock In: Proceedings of the 2022 ACM on Asia Conference on Computer and
  Communications Security (2022).
\newblock \doi{10.1145/3488932.3497768}

\bibitem{gymoai}
{The Farama Foundation}: Announcing the {Farama Foundation}: The future of open
  source reinforcement learning (2022).
\newblock \urlprefix\url{https://farama.org/Announcing-The-Farama-Foundation}

\bibitem{pesidious}
Vaya, C., Sen, B.: Pesidious: Malware mutation using reinforcement learning and
  generative adversarial networks (2020).
\newblock \urlprefix\url{https://github.com/CyberForce/Pesidious}.
\newblock Accessed: 2026-01-26

\bibitem{virustotal}
{VirusTotal}: {VirusTotal} -- online malware analysis service (2026).
\newblock \urlprefix\url{https://www.virustotal.com}.
\newblock Accessed: 2026-03-09

\end{thebibliography}

\end{document}